\documentclass[12pt,a4paper]{article}%
\usepackage{amsfonts}
\usepackage{amssymb}
\usepackage{amsmath}
\usepackage{natbib}
\usepackage{scalefnt}%
\usepackage{booktabs}  
\usepackage{graphicx}
\usepackage{float}
\usepackage[hidelinks,pdftex]{hyperref}

\usepackage{afterpage}
\usepackage{geometry}

\usepackage{subcaption}
\usepackage{bm}
\usepackage{bbm} 

\usepackage[table,dvipsnames]{xcolor}  
\usepackage{textcomp}
\usepackage{titling}
\usepackage{url}

\usepackage{wrapfig}
\usepackage{lscape}
\usepackage{rotating}
\usepackage{epstopdf}
\usepackage{pdfpages}

\usepackage{tikz}

\definecolor{UF}{HTML}{D95319}
\definecolor{Actual}{HTML}{0072BD}
\definecolor{Q1}{HTML}{EDB120}
\definecolor{Q2}{HTML}{7E2F8E}
\definecolor{Q3}{HTML}{77AC30}
\definecolor{Q4}{HTML}{4DBEEE}
\newcommand{\emptyauthor}{\global\let\@author\@empty}
\hypersetup{colorlinks=true,urlcolor=blue, citecolor=blue, linkcolor= black}

\providecommand{\U}[1]{\protect\rule{.1in}{.1in}}

\RequirePackage{natbib}

\usepackage{doi-frbc}

\usepackage{authblk}

\begin{document}

\title{Monthly GDP Growth Estimates for the U.S. States\thanks{We thank conference and seminar participants at the Midwest Econometrics Group, the SGPE conference, University of Strathclyde, CFE, ESCoE Economic Measurement, the Cleveland Fed, University of Salzburg, Athens University of Economics and Business, and the Federal Reserve System Committee on Regional Analysis for helpful comments. The views expressed herein are those of the authors and not necessarily those of the Federal Reserve Bank of Cleveland or the Federal Reserve System. }}

\author{Gary Koop\thanks{University of Strathclyde. 
        Email: \href{gary.koop@strath.ac.uk}{gary.koop@strath.ac.uk} },  
Stuart McIntyre\thanks{University of Strathclyde. 
        Email: \href{s.mcintyre@strath.ac.uk}{s.mcintyre@strath.ac.uk}}, 
James Mitchell\thanks{Federal Reserve Bank of Cleveland. 
        Email: \href{james.mitchell@clev.frb.org}{james.mitchell@clev.frb.org} }, and 
Aristeidis Raftapostolos\thanks{King's College London. 
        Email: \href{aristeidis.1.raftapostolos@kcl.ac.uk}{aristeidis.1.raftapostolos@kcl.ac.uk}} } 

\maketitle

\vspace{-1ex}
\begin{center}
    \date{} 
    
\end{center}

\begin{abstract}
    \noindent
    This paper develops a mixed frequency vector autoregressive (MF-VAR) model to produce nowcasts and historical estimates of monthly real state-level GDP for the 50 U.S. states, plus Washington DC, from 1964 through the present day. The MF-VAR model incorporates state and U.S. data at the monthly, quarterly, and annual frequencies. Temporal and cross-sectional constraints are imposed to ensure that the monthly state-level estimates are consistent with official estimates of quarterly GDP at the U.S.  and state-levels. We illustrate the utility of the historical estimates in better understanding state business cycles and cross-state dependencies. We show how the model produces accurate nowcasts of state GDP three months ahead of the BEA's quarterly estimates, after conditioning on the latest estimates of U.S. GDP. 
 \bigskip
    \noindent
    
\textit{Keywords: } State economies; Monthly GDP; Mixed frequency; Regional nowcasting; Vector Autoregression; Business cycles; Cross-state heterogeneity; Bayesian analysis\\

\noindent
JEL Classification: C32, C55, C80, E01, R11

\end{abstract}

\newpage

\section{Introduction}

Real gross domestic product (GDP) remains, in the words of the NBER's Business Cycle Dating Committee, the ``single best measure of aggregate economic activity.'' While national statistical offices have made progress both in speeding up and increasing the frequency of their estimates of national-level GDP, regional or state-level GDP estimates remain subject to longer publication lags and to shorter historical samples. To give an example, and the one that motivates this paper, in the U.S. the Bureau of Economic Analysis (BEA)'s quarterly (real) gross state product estimates (GSP, hereafter referred to simply as state GDP)  are available back to 2005; prior to which they are only available at the annual frequency. Moreover, state GDP estimates are published with greater delay than equivalent national data. Typically, state GDP estimates are now released by the BEA three months after the end of the calendar quarter. 

These deficiencies in the measurement of state GDP impede both real-time tracking of regional and state economies and historical business cycle analysis seeking to understand if and how these economies differ in their dynamics and responsiveness to shocks. Both exercises are important, not least for policymakers setting monetary and fiscal policy. Even when interest resides with the aggregate economy, understanding the differential behavior of state or regional economies helps explain the workings of the aggregate economy; for example, see \cite{Carlino1999}. Cross-sectional data are also used to help identify structural macroeconomic relationships, such as the Phillips curve; for example, see \cite{NakamuraQJE}. As a result of the lack of higher-frequency and long samples of state GDP data, economists seeking to study state business cycles have often had to compromise in terms of the data that they use.\footnote{Previous studies impeded by the lack of higher-frequency state GDP data include the following. \cite{Carlino1999} explain that they use state-level quarterly real personal income data in the absence of higher-frequency state GDP data. \cite{Gonzalez2019} estimates output gaps across the U.S. states, but only using quarterly state GDP data from 2005. \cite{Kouparitsas2002} quantifies the importance of cross-state spillovers and common shocks using personal income data, noting the lack of higher-frequency state GDP data forces this data choice. \cite{Owyang2005} identify business cycle phases in U.S. states using the monthly state-level coincident indexes of \cite{crone2005consistent} because the state GDP data is available at too low a frequency, too short a historical sample, and with too much of a delay.}

The absence of higher frequency state GDP data with historical coverage sufficient for meaningful business cycle analysis has not, however, stopped economists from exploiting the rich array of higher frequency indicator data that are available at the state-level. These indicators include labor market data, personal income data, and financial market data. A leading example is produced by the Federal Reserve Bank of Philadelphia. Each month they publish a timely monthly coincident economic indicator for each state in the U.S..\footnote{See \tiny{\url{https://www.philadelphiafed.org/surveys-and-data/regional-economic-analysis/state-coincident-indexes}}.} This involves using a dynamic factor model, as set out in \cite{crone2005consistent}, to combine four monthly and quarterly indicators of state-level economic activity. More recently, \cite{Baumeister} developed weekly indicators of state-level economic activity. They also use a dynamic factor model but combine it with a larger and richer set of weekly, monthly, and quarterly indicators. Consideration of weekly data, including non-standard variables (such as mobility measures), is shown to be especially helpful when monitoring the economic performance of state economies during the COVID-19 pandemic. 

But while each of these indices offers a useful higher-frequency and more timely indicator of state-level economic activity, neither is intended to provide direct higher-frequency estimates of state GDP itself, even though we might expect the indicators consulted to inform on its path. This is the challenge taken up in this paper: combining and reconciling different mixed-frequency sources of data to produce more timely and higher-frequency estimates of state GDP.\footnote{We note that similar issues are faced in other countries, as emphasized by \cite{Stock2005}: “...an important practical challenge facing regional economists is combining ...different sources of data to provide a timely and accurate measure of regional economic activity.”} We focus on the production of monthly state GDP estimates. But we note that our model could, albeit at considerable further computational expense, be extended to produce weekly state GDP estimates too. Arguably, the estimation of state GDP at the monthly frequency is sufficient for regional and state business cycle analysis, certainly outside of, for example, pandemic periods when important events were unfolding intra-month. 

We develop a model to produce monthly estimates of real state GDP that conditions not only on available monthly indicators, such as the labor market variables used by the Philadelphia Fed, but also on the official but lower-frequency estimates of state- and U.S.-level GDP produced by the BEA. To be consistent with these official data, monthly estimates of state GDP must respect both temporal and cross-sectional aggregation constraints. First, the monthly state GDP data must be temporally consistent with the BEA's own estimates of quarterly (annual prior to 2005) state GDP, once published. Second, the state GDP data need to be consistent with BEA estimates of quarterly U.S. GDP data: state GDP cross-sectionally has to aggregate to U.S. GDP. The need to impose this cross-sectional constraint explains why, extending \cite{crone2005consistent} and \cite{Baumeister}, we estimate our model jointly across the 50 states (plus Washington, DC), rather than separately, state-by-state. A further advantage of joint estimation is it means that we can condition state GDP nowcasts on the more timely U.S. GDP estimates. The US GDP data, in effect, can be used to help guide the state GDP data when apportioning the more timely US data across the states. Econometrically, both sets of constraints are incorporated into a Bayesian mixed-frequency vector autoregressive (MF-VAR) model that accommodates cross-state spillovers and dependencies. 

The proposed MF-VAR model, given interest in estimating and interpolating state GDP itself, is deliberately specified as a state space or parameter-driven model. A leading alternative class of MF-VAR model, that does not rely on latent processes, was developed by \cite{ghysels2016macroeconomics}. Such observation-driven mixed data-sampling frequency (MIDAS) models, while not permitting the estimation of monthly state GDP, which is our focus, are useful when interest rests with impulse response analysis. Mixed data-sampling models have also been used successfully in many applications, such as forecasting U.S. state government revenues and expenditures; see \cite{ghysels2022real}.

Our MF-VAR is fundamentally a very large one, involving over $50$ equations (for each state) and a three-way frequency mismatch that changes over time (that is, state level GDP is initially annual, then quarterly, and then other variables are either quarterly or monthly). Accordingly, we use the horseshoe prior of \cite{carvalho2010horseshoe}. This is a Bayesian shrinkage prior popular in the machine learning literature which selects, in an automatic fashion, the important coefficients, and shrinks the remainder to zero ensuring parsimony. Bayesian estimation with VARs is typically done using Markov chain Monte Carlo (MCMC) methods. The computational burden of conventional MCMC methods applied to MF-VARs becomes prohibitive in large models such as the one we have. To address this, we develop a computationally fast approximate MCMC algorithm which allows us to carry out estimation and nowcasting. The model thus enables a richer characterization of the higher-frequency effects of shocks on the U.S. states. A final advantage of producing higher-frequency estimates of state GDP itself, rather than estimates of a (latent) coincident index, is that these estimates can be evaluated. That is, the model's monthly state GDP estimates, once aggregated to the quarterly frequency, can be compared (and evaluated) against BEA estimates of quarterly state GDP once, with a greater lag, they are published. In a real time application nowcasting state GDP, we show that accurate nowcasts can be obtained via our MF-VAR once two months of within-quarter information are known and the model is, in effect, conditioning the state GDP nowcasts on reliable estimates of U.S. GDP. The state GDP nowcasts are more accurate than those from a MF-VAR model that neither allows for cross-state dependencies nor imposes the cross-sectional constraint disciplining the estimates of state GDP so that they are consistent with (that is, aggregate to) known estimates of U.S. GDP. These state GDP nowcasts are available three months ahead of BEA data.

The plan for the remainder of this paper is as follows. Section 2 sets out the proposed cross-state MF-VAR model and explains the Bayesian estimator.  Section 3 considers the application that uses the MF-VAR to produce historical monthly estimates of state GDP from 1964 through 2024. We illustrate the use of these new data for state business cycle analysis and the study of state connectedness. Then we show how the MF-VAR model can be used to produce accurate nowcasts of state GDP that are available at least three months ahead of BEA data. The production and dissemination of timely higher-frequency estimates of state GDP will be useful for decision makers tracking the evolution of state economies. Section 4 concludes. Online appendices comprise a data Appendix (Appendix \ref{DataAppendix}), a technical Appendix (Appendix \ref{TechAppendix}) detailing the estimation algorithm, and an empirical Appendix (Appendix \ref{EmpiricsAppendix}) containing supplementary empirical results.

\section{Econometric Methods}

\subsection{Mixed-frequency Models with Large Data Sets}

Dynamic factor models (DFMs) have long been used when working with large macroeconomic data sets. But since the pioneering paper of \cite{BGR2010}, large VARs have also enjoyed a surge in popularity. Bayesian methods are typically used, given the need for prior shrinkage to overcome the proliferation of parameters problem that occurs in large VARs.  The success of large VARs is partly due to the fact that, beginning with \cite{BGR2010}, they have been found to forecast better across a range of macroeconomic data sets. But it is also because the unrestricted nature of the large VAR aids in interpretation and structural analysis. For instance, with a large VAR it is easy to see the role that individual variables play in informing forecasts in a manner which is difficult when using factors which compress the information in many variables together into a small number of factors. In our approach, where we want to jointly model all the U.S. states (alongside the U.S. as a whole) in a single model, so as to investigate inter-relationships between them, the large VAR approach seems a natural choice.  This division of the literature into DFM and VARs has also occurred when using mixed-frequency data and, although there have been many successful mixed-frequency DFMs, in this paper we use MF-VARs.  

MF-VARs have enjoyed great popularity in policy circles since they can provide timely, high-frequency nowcasts of low-frequency variables such as GDP growth which are released with a delay. For instance, \cite{schorfheide2015real}, \cite{bbj2019}, and \cite{mccracken2019real} are influential papers associated with the Federal Reserve Banks of Minneapolis, Chicago, and St. Louis, respectively. A common set-up is to nowcast a quarterly variable (for example, GDP growth in the U.S.) using several monthly variables. Nowcasting state GDP growth is more of a challenge, since there are $51$ variables to be nowcast and the frequency mismatch is more complicated (that is, we have a three-way frequency mismatch involving annual, quarterly and monthly variables) and this mismatch changes over time. Specifically, state GDP growth is available at an annual frequency through 2004, before the BEA started producing quarterly estimates from 2005. U.S. GDP growth is available quarterly throughout the sample, and we include many monthly indicator variables in the model to provide a better impression of within-quarter business cycle dynamics. To complicate the data landscape further, the variables all have a range of release delays and are available over different historical sample periods. 

In light of these data features, in this paper we work with MF-VARs that are much larger than is conventional, and involve many more, and more complicated, latent states. This raises challenges in terms of over-parameterization concerns and the computational burden. We develop a Bayesian modeling framework which overcomes these challenges, and use it to produce monthly estimates of state GDP. Importantly our estimates are consistent with the BEA's estimates of both quarterly state- and U.S.-level GDP. Consistency is achieved, following \cite{KOOP2022}, by imposing inter-temporal and cross-sectional aggregation constraints within our MF-VAR that link published data with the model-based estimates.

\subsection{Notation and Data Observability}

 All our variables enter our models in growth rates and are denoted by $y$.\footnote{These growth rates are exact, not log differenced. Results using log differences are very similar; see the online Appendix \ref{EmpiricsAppendix}. As explained in \cite{kmmp1}, when modeling in log differences the aggregation constraints presented below require minor modification.}  We use a notational convention where subscripts $a,q,m$ denote annual, quarterly, and monthly frequencies, respectively. The first subscript on any variable denotes the frequency at which that variable is observed.\footnote{For convenience, we denote state-level variables with an $a$ even though they are observed quarterly from 2005 onwards.} The second subscript, when used, indicates that the variable is being modeled at a different frequency to the observed one. For example, prior to 2005 state GDP growth was observed only at the annual frequency and so, to refer to latent monthly state GDP growth, we use the subscripts $a,m$. If the first two subscripts are the same we suppress one of them (for example, employment growth is observed at the monthly frequency and, thus, we simply use subscript $m$ instead of $m,m$). The third subscript $t=1, \ldots, T$ denotes time at the monthly frequency. Superscripts $US$ and $ s=1, \ldots S$ distinguish between variables for U.S. as a whole and the individual states. With 50 states plus DC, $S=51$.
 
 Our model involves the following variables: 
 \begin{itemize}
\item $ y_{m, t}^{US}$ is a vector of monthly macroeconomic variables for the U.S. which are always observed (e.g., employment growth). 
\item $ y_{m,q, t}^{US}$ is a vector of quarterly observations on the preceding U.S. macroeconomic variables. It is constructed from $ y_{m, t}^{US}$ and, thus, is always observed. 
\item $y_{q,m,t}^{US}$ is a vector of monthly variables for the U.S. for the variables which are observed at the quarterly frequency. It is never observed at the monthly frequency. The first variable in this vector is GDP growth and, when we wish to isolate this variable, we will include an additional * superscript. 
\item $y_{q,t}^{US}$ is a vector of quarterly variables for the U.S.. It is observed for months 3, 6, 9 and 12, but not in other months.
\item $ y_{a,m,t}^{s}$ is monthly GDP growth in state s. It is never observed.
\item $ y_{a,q,t}^{s}$ is quarterly GDP growth in state s. Prior to 2005 it is never observed. From 2005 onwards it is observed for months 3, 6, 9 and 12, but not in other months.
\item $ y_{a,t}^{s}$ is annual GDP growth in state $s$. Prior to 2005 it is observed for month 12 of every year. 
\end{itemize}

If the $t$ subscript is suppressed, it denotes the vector of all observations on a variable. Superscript $s$ denotes the vector containing quantities for all $51$ states. 

Given that they should help inform the path of (latent) monthly state-level GDP, we also consider a set of observed monthly state-level indicators.  These state-level indicators, chosen as discussed below to help track intra-year and intra-quarter movements in state GDP, are included as exogenous monthly variables in the MF-VAR. As a result, for notational ease, we do not explicitly distinguish them in the equations below. These state-level exogenous variables are included in a state-specific manner to ensure parsimony. That is, an exogenous variable for state $i$ only appears in the equation for state $i$.\footnote{A minority of these exogenous state-level variables are in fact observed only at the quarterly frequency. To avoid further increasing the size of an already large VAR model, by adding in more state equations so that we can model the underlying latent monthly state-level variable, we simply assume constant growth across the three months of a given quarter.}  

\subsection{The MF-VAR with Inter-temporal and Cross-sectional Constraints}

We write the MF-VAR as:
\begin{equation}
Ay_{t}= B_0 + B_{1}y{}_{t-1}+ \cdots +B_{p}y{}_{t-p}+ \epsilon_{t},\epsilon_{t}\sim N(0,\Sigma),\label{VAR}
\end{equation}
for $t=1,\ldots,T$ where $y_t$ is a vector of $N\times1$ monthly dependent variables ordered as $y_{m,t}^{US}$ (observed monthly variables), $y_{q,m,t}^{US}$ (unobserved monthly U.S. variables), then $y_{a,m,t}$ (unobserved monthly state-level GDP growth). $A$ is a lower triangular matrix, with ones on the diagonal, $B_{i}$ for $i=1, \ldots, p$ are the VAR coefficient matrices, and $\Sigma$ is a diagonal matrix with diagonal elements denoted by $\sigma_{i}^2$ for $i=1,\ldots,N$.\footnote{In our empirical work we add intercepts and state-specific exogenous variables to this specification. We choose a relatively long lag length of $5$ and trust our horseshoe prior (described below) to shrink extraneous coefficients to zero.}

Writing the MF-VAR in structural form with $\Sigma$ being diagonal greatly reduces the computational burden since it allows
for equation-by-equation estimation of the model (see, e.g., \cite{carriero2019large}) and does not restrict the reduced form error covariance.  
Each equation of our model can be written as:
\begin{equation}\label{eq3}
    y_{i,t} = w_{i,t} \alpha_{i} + x_{i,t} \beta_{i} + \varepsilon_{i,t}
\end{equation}
where $\varepsilon_{i,t} \sim N (0, \sigma^2_{i})$, $ w_{i,t} = \begin{pmatrix}
    -y_{i,t} & \ldots & -y_{i-1,t} \end{pmatrix}'$, $ x_{i,t} = \begin{pmatrix}
    1 & y_{t-1} & \ldots & y_{t-p} \end{pmatrix}'$, $\beta_{i}$ and $\alpha_{i}$ are the rows of the VAR coefficients and $A$ associated with the $i$-th equation.
   Below we use notation where $X_{i,t} = \begin{pmatrix} w_{i,t}$ $x_{i,t} \end{pmatrix}$ and $\theta_{i} = \begin{pmatrix}
    \alpha_{i}^\prime & \beta_{i}^\prime \end{pmatrix}'$, where $\theta_{i}$ has dimension $k_i = Np+i$.
   
The MF-VAR is a state space model, where (\ref{eq3}) provides us with the state equations for the partially unobserved $y_t$. The measurement equations in the state space model specify the observability conditions for every variable and link the observed low frequency variables to their unobserved high frequency counterparts via inter-temporal restrictions. In our model, we have a three-way frequency mis-match involving variables which are observed at the monthly, quarterly, and annual frequencies. Different inter-temporal restrictions apply for the various frequency mis-matches.

Recall that quarterly state GDP growth is observed after 2005. The inter-temporal restriction linking this to its unobserved monthly state counterpart post-2005 can be shown to be (see \cite{mariano2003new} and \cite{kmmp1}): 

\begin{equation}
y_{a,q,t}^{s} =\frac{1}{3}y_{a,m, t}^{s}+\frac{2}{3}y_{a,m, t-1}^{s}+y_{a,m, t-2}^{s}%
+\frac{2}{3}y_{a,m, t-3}^{s}+\frac{1}{3}y_{a,m, t-4}^{s} 
\label{q_to_m_states}
\end{equation}
for $s=1,..,S$. An inter-temporal restriction of the same form links monthly U.S. GDP growth to its observed quarterly value:
\begin{equation}
y_{q,t}^{US,*} =\frac{1}{3}y_{q,m, t}^{US,*}+\frac{2}{3}y_{q,m, t-1}^{US,*}+y_{q,m, t-2}^{US,*}%
+\frac{2}{3}y_{q,m, t-3}^{US,*}+\frac{1}{3}y_{q,m, t-4}^{US,*} 
\label{q_to_m_US}
\end{equation}

Prior to 2005, state GDP growth was only observed annually and the inter-temporal restriction linking the observed quantity to the desired monthly quantity is:

\begin{align}
y_{a,t}^{s} &= \frac{1}{12}y_{a,m,t}^{s} + \frac{2}{12}y_{a,m,t-1}^{s} + \frac{3}{12}y_{a,m,t-2}^{s} 
+ \frac{4}{12}y_{a,m,t-3}^{s} + \frac{5}{12}y_{a,m,t-4}^{s} \nonumber \\
&\quad + \frac{6}{12}y_{a,m,t-5}^{s} + \frac{7}{12}y_{a,m,t-6}^{s} + \frac{8}{12}y_{a,m,t-7}^{s} 
+ \frac{9}{12}y_{a,m,t-8}^{s} + \frac{10}{12}y_{a,m,t-9}^{s} \nonumber \\
&\quad + \frac{11}{12}y_{a,m,t-10}^{s} + \frac{12}{12}y_{a,m,t-11}^{s} + \frac{11}{12}y_{a,m,t-12}^{s} 
+ \frac{10}{12}y_{a,m,t-13}^{s} \nonumber \\
&\quad + \frac{9}{12}y_{a,m,t-14}^{s} + \frac{8}{12}y_{a,m,t-15}^{s} + \frac{7}{12}y_{a,m,t-16}^{s} 
+ \frac{6}{12}y_{a,m,t-17}^{s} \nonumber \\
&\quad + \frac{5}{12}y_{a,m,t-18}^{s} + \frac{4}{12}y_{a,m,t-19}^{s} + \frac{3}{12}y_{a,m,t-20}^{s} 
+ \frac{2}{12}y_{a,m,t-21}^{s} + \frac{1}{12}y_{a,m,t-22}^{s}, 
\label{a_to_m}
\end{align}
for $s=1 \ldots S$. 

Other measurement equations can be obtained through cross-sectional restrictions which arise from the fact that U.S. GDP is the sum of GDP for the individual states. We apply these at both the monthly and quarterly frequency.\footnote{The monthly version of the cross-sectional restriction is not really a measurement equation since it involves only latent states. Nevertheless, it does help inform our monthly estimates of state GDP due to the U.S. intertemporal restriction. That is, unobserved monthly state GDP adds up to unobserved monthly U.S. GDP, but the latter is constrained to add up to observed quarterly U.S. GDP.}  For exact growth rates, the cross-sectional restriction at the quarterly frequency can be shown to be (see \cite{kmmp1}):

\begin{equation}
y_{q,t}^{US,*} = 
\sum_{s=1}^{S} w_{t}^{s} y_{a,q,t}^{s} + \epsilon^{cs}_t,
\label{cross_sec}
\end{equation}
where $w_{t}^{ s}$ is the share of state-level output in aggregate output in
quarter $t$ and $\epsilon^{cs}_{t}\sim N(0,\sigma_{cs}^{2})$.\footnote{\textcolor{black}{In the absence of BEA data for monthly state-level and U.S. GDP in levels (in dollars), we proxy these weights with the sample average of the share of each state's real GDP in U.S. GDP. Specially, from 1977 we use the real state GDP data available at \href{https://apps.bea.gov/itable/?ReqID=70\&step=1\#eyJhcHBpZCI6NzAsInN0ZXBzIjpbMSwyNCwyOSwyNSwzMSwyNiwyNywzMF0sImRhdGEiOltbIlRhYmxlSWQiLCI1MjIiXSxbIkNsYXNzaWZpY2F0aW9uIiwiU0lDIl0sWyJNYWpvcl9BcmVhIiwiMCJdLFsiU3RhdGUiLFsiMCJdXSxbIkFyZWEiLFsiWFgiXV0sWyJTdGF0aXN0aWMiLFsiMSJdXSxbIlVuaXRfb2ZfbWVhc3VyZSIsIkxldmVscyJdLFsiWWVhciIsWyItMSJdXSxbIlllYXJCZWdpbiIsIi0xIl0sWyJZZWFyX0VuZCIsIi0xIl1dfQ==}{BEA} and from 1963 to 1976 use the current-price state-level GDP series deflated by the U.S. CPI, as described in the data Appendix \ref{DataAppendix}. This does mean that we ignore temporal variation in the shares of each state in U.S. GDP; relaxing this to allow the weights to update each year does not affect our results.}} 
The restriction at the monthly frequency is the same except that $y_{q,t}^{US,*}$ is replaced by $y_{q,m.t}^{US,*}$ and $y_{a,m,t}^{s}$. 

Note that the quarterly cross-sectional restriction becomes less useful in 2005 when quarterly state-level data becomes available. However, it is applied in our econometric model as it is of some use when nowcasting due to the difference in release delays for U.S. and state-level GDP data. For both cross-sectional restrictions we proxy $w_{t}^{s}$ by the observed annual shares, noting that we should expect to see little
within-year variation in these weights. Note that we add an error to these cross-sectional restrictions since state GDP need not sum to exactly to U.S. GDP. The reason for this is that U.S. GDP equals the sum of state GDP and overseas activity, comprising, notably, military economic activity. Since overseas activity has quite distinct time-series properties from the other variables in our model, we chose not to include it, thereby introducing a small error into the cross-sectional constraint, captured by $\epsilon^{cs}_t$. 

There is also a known discontinuity in state-level GDP in 1997. Prior to 1997, the BEA measured state GDP such that these data cross-sectionally aggregated to U.S. gross domestic income (GDI), whereas thereafter they aggregate to U.S. GDP (on the expenditure-side).\footnote{See \url{https://www.bea.gov/cautionary-note-about-annual-gdp-state-discontinuity}} Since GDP and GDI do not equate exactly in practice this could lead to a small error in the cross-sectional restriction.\footnote{By interpolating quarterly state GDP prior to 1997 with respect to GDP, rather than GDI, we produce historical time-series for state GDP less exposed to this data-discontinuity critique, since the quarterly state GDP estimates are always reconciled by U.S. GDP.} A final rationale for allowing for a stochastic error is that, in real time, when mixing data vintages for U.S. and state GDP measurement errors contribute to state GDP not always adding up to U.S. GDP.

In summary, our MF-VAR is a state space model with state equations given by (\ref{eq3}) and measurement equations given by (\ref{q_to_m_states}), (\ref{q_to_m_US}), (\ref{a_to_m}), and the quarterly and monthly versions of (\ref{cross_sec}).

\subsection{The Prior for the MF-VAR}

The MF-VARs estimated in this paper are all of dimension in excess of $50$. When working with VARs of this dimension, there is a strong need for prior shrinkage and many alternatives have been proposed in the literature including forms inspired by the classic Minnesota prior such as \cite{BGR2010}, variable selection priors such as \cite{k2013}, and global-local shrinkage priors such as \cite{kh2020}. With MF-VARs the need for prior shrinkage becomes even more important due to the additional need to estimate the high frequency values of the variables which are only observed at a low frequency. In this paper, we adopt the horseshoe prior proposed by \cite{carvalho2010horseshoe} using the implementation of \cite{korobilis2022new}. The horseshoe prior belongs to the family of global-local shrinkage priors. We implement the horseshoe prior one equation at a time. That is, each equation in the MF-VAR has its own prior allowing for a different degree of shrinkage in each equation. For equation $i$, for $i=1 \ldots N$, we have a global shrinkage prior parameter, $\tau_i$, and a local shrinkage parameter, $\lambda_{ij}$, which is specific to the $j^{th}$ coefficient. The horseshoe prior has properties which are often found advantageous in sparse models. It aggressively penalizes small coefficients, but applies minimal shrinkage to large coefficients. Thus, the noise provided by large numbers of irrelevant coefficients in the MF-VAR is largely removed and the signal provided by the few non-zero coefficients is more precisely estimated in a data-based fashion. Since the horseshoe prior does not require a choice of prior hyperparameters and we follow the specification in \cite{korobilis2022new} exactly, we do not provide further details here. 

The remaining parameter is $\sigma_{cs}^{2}$, which is the variance of the error in the cross-sectional restriction. This restriction is of importance in obtaining accurate estimates of monthly state GDP, since it the main avenue in which newly released GDP growth figures for the U.S. as a whole spill over into estimates for the individual states. If $\sigma_{cs}^{2}$ is a small number, then this link between the U.S. and the individual states is strengthened. Larger numbers weaken this link. Of course, this parameter is estimated from the data, but its prior can influence the estimate. We assume:
\begin{equation}
    \sigma_{cs}^2 \sim IG \left( \underline{ \nu_{\sigma}} , \underline{S_{\sigma}} \right),
\end{equation}%
with $\nu_{\sigma} = 10$ and $\underline{S_{\sigma}}=0.01$. These choices imply that the prior mean of $\sigma^2$ is $0.0001$, which is roughly consistent with the magnitude of our growth rates data, but relatively non-informative. We use the same prior for the monthly and quarterly cross-sectional restrictions.

\subsection{Posterior Inference in the MF-VAR Using a Computationally Efficient Approximate MCMC Algorithm}

Bayesian inference in MF-VARs can be undertaken using MCMC methods. For the conventional MF-VAR with a single frequency mismatch and no cross-sectional restriction, the algorithm of \cite{schorfheide2015real} is commonly used. \cite{kmmp1} extend this algorithm to an MF-VAR with a cross-sectional restriction. \cite{kmmp2} further extend the algorithm for a case where the frequency mismatch changes over time. Small adaptations of these algorithms are required to handle the three-way frequency mismatch involved in our U.S. state-level application. We considered such adaptions, but found that the MCMC algorithm is simply too slow to carry out extensive empirical work or a real time forecasting exercise.\footnote{The empirical work in this paper would take months or years even on a high-quality personal computer.} 

The fact that MCMC methods are not scaleable to models with high-dimensional parameter spaces is well-known to Bayesian econometricians. This lack of scalability is a particular problem in large VARs and has led to the use of approximate methods. For instance, \cite{gkp2020} develop Variational Bayesian (VB) methods for the MF-VAR. VB methods are computationally efficient but are an approximate method. It is well-known that, although they provide accurate approximations to posterior means, they tend to under-estimate posterior variances. In a forecasting exercise this leads to an under-estimation of predictive variances, see \cite{gkp2021}. In this paper, we propose an approximate method specifically designed for our MF-VAR with a three-way frequency mismatch which does not rely on VB methods or similar. It is MCMC based, thereby allowing for full exploration of the high-dimensional posterior and accurate reflection of posterior and predictive uncertainty.

MCMC algorithms for drawing the parameters of an MF-VAR (conditional on draws of latent states for the low frequency variables) are now standard. Conditional on the latent states, the model reduces to a VAR and Bayesian methods for VARs are available. For instance, Gibbs samplers for Bayesian VARs with several global-local shrinkage including the horseshoe are given in \cite{gkp2021}. Hence, we will not describe them here nor explicitly list the VAR parameters as conditioning arguments in this sub-section.  Instead, we will develop methods for drawing monthly GDP growth (conditional on draws of the parameters). As noted above, our model is a Normal linear state space model and standard methods exist for drawing latent states in such models. However, the exact MCMC algorithm is very computationally burdensome in MF-VARs of dimension exceeding $50$ such as the ones used in this paper. In practice, we have found the main computational bottleneck lies in the parts of the model involving the annual-monthly frequency mismatch and, in particular, the fact that its inter-temporal restriction, given in (\ref{a_to_m}), involves over 20 lags. Accordingly, we develop an approximate MCMC algorithm which avoids the use of the annual-monthly inter-temporal restriction. The idea underlying our algorithm is that it is much faster to draw from two separate algorithms, one involving a quarterly-monthly frequency mismatch and the other involving a quarterly-annual mismatch. In the remainder of this sub-section we show precisely how this can be achieved. 

Consider an MCMC algorithm which produces draws of state-level monthly and quarterly GDP growth rates, given U.S. data and annual state level data. The posterior in this case is 
$p(y_{a,m}^{S},y_{a,q}^{S}|y_{a}^{S},y_{q}^{US},y_{m}^{US})$. A simple rule of probability implies:
\begin{equation}
p(y_{a,m}^{S},y_{a,q}^{S}|y_{a}^{S},y_{q}^{US},y_{m}^{US})=p(y_{a,m}^{S}|y_{a,q}^{S},y_{a}^{S},y_{q}^{US},y_{m}^{US})
p(y_{a,q}^{S}|y_{a}^{S},y_{q}^{US},y_{m}^{US})
\label{2step}
\end{equation}

The first of the term on the right-hand-side of (\ref{2step}) can be simplified to $p(y_{a,m}^{S}|y_{a,q}^{S},\allowbreak y_{q}^{US},y_{m}^{US})$ since, conditional on knowing the quarterly growth rates ($y_{a,q}^{S}$), the annual growth rates $y_{a}^{S}$ provide no additional information. But this is the posterior that arises in a conventional MF-VAR involving a quarterly/monthly frequency mismatch, no annual quantities appear in it at all. The MCMC algorithm for such a posterior is standard (see, for example, \cite{schorfheide2015real}).

The second term on the right-hand side of (\ref{2step}) does involve quarterly, monthly, and annual variables and thus will involve the annual-monthly inter-temporal restriction. But consider the posterior density $p(y_{a,q}^{S}|y_{a}^{S},y_{q}^{US},y_{m,q}^{US})$, which is the same as this second term except for the fact that $y_{m}^{US}$ is replaced by $y_{m,q}^{US}$ (i.e., monthly variables such as employment growth rates have been replaced by quarterly growth rates). The MCMC algorithm for this posterior is also a conventional one for an MF-VAR, involving an annual/quarterly frequency mis-match as in, e.g., \cite{kmmp1}.

This reasoning suggests the following strategy: use two conventional MF-VARs, one involving a quarterly/monthly frequency mismatch and one involving an annual/quarterly frequency mismatch. Draws of $y_{a,q}^{S}$ produced by the algorithm of the second MF-VAR are then conditioned on in the first MF-VAR. We have found this strategy to be much more computationally efficient than drawing directly from the algorithm for the three way frequency mismatch, since it avoids dealing with the computational bottleneck caused by having annual and monthly variables in the same model. 

This strategy is an approximate one, since the algorithm which produces quarterly draws of state GDP growth is conditional on quarterly versions of monthly variables instead of monthly. But the loss of information is likely to be small (i.e., to produce quarterly estimates of state GDP knowing the U.S. quarterly quantities will be useful, but knowing them at the monthly frequency is likely to provide only minimal improvements). We stress that this approximation is only used in one part of the algorithm. The important part of the algorithm is $p(y_{a,m}^{S}|y_{a,q}^{S},y_{q}^{US},y_{m}^{US})$ which produces the draws of monthly GDP growth for each state. This does condition on the monthly data and, thus, our monthly estimates of state GDP growth do reflect the information contained in monthly predictors. If we further consider that this approximate algorithm is only used for producing pre-2005 draws of quarterly state GDP growth (since subsequently quarterly state level data are available), the argument in favor of gaining large computational benefits by using this approximate algorithm is strengthened.

\section{Empirical Application}

\subsection{State and U.S. Data}

Our mixed-frequency real time dataset consists of quarterly U.S. real GDP, annual and quarterly state-level real GDP, plus 13 U.S. macroeconomic indicators and 6 state-level variables available at the monthly or quarterly frequency. The online Data Appendix (\ref{DataAppendix}) provides a list of all the variables and their transformations. For each category, the variables are ordered as listed in this Appendix. The states enter the MF-VAR in alphabetical order. Here we provide a brief description of these variables and a motivation for their inclusion. 

We start with the real GDP data. For the U.S., these data are available on a quarterly basis back to the beginning of our sample. We set the beginning of this sample as 1964 (in growth rates), since 1963 is the first year of available data from the BEA for state GDP. Prior to 1977, these annual state GDP data are available only in nominal terms, so in the absence of state-level price data we follow \cite{del2002asymmetric} and deflate by the U.S. GDP deflator. From 2005, we use the quarterly real state-level GDP data published by the BEA. Our historical (in-sample) analysis uses latest vintage estimates of U.S. and state-level GDP (defined as the June 2024 vintage). However, in the out-of-sample exercise, designed to test the ability of our model to produce accurate but more timely monthly state GDP, we use real time data vintages.

Given our focus on producing higher-frequency and historical state GDP data consistent with official data, the other macroeconomic indicators we use are also sourced from the BEA and the Bureau of Labor Statistics, with additional indicators and real-time vintages (discussed below) obtained from Bloomberg, and the ALFRED and FRED-SD (\cite{BOKUN2022}) real-time databases maintained by the Federal Reserve Bank of St. Louis. 

As our starting point is the MF-VAR model used by \cite{schorfheide2015real} to estimate monthly GDP at the U.S.-level, we consider the same set of quarterly and monthly U.S. observed variables. \cite{schorfheide2015real} consider three quarterly variables, GDP, fixed investment, and government expenditure, and 8 monthly variables: industrial production, personal consumption expenditure, hours worked, the unemployment rate, CPI inflation, the $S\&P500$, the effective federal funds rate, and the ten-year Treasury rate. We then add the following three monthly indicators at the U.S.-level: employment (total nonfarm), real personal income, and crude oil prices. We added personal income as \cite{arias2016metro} and \cite{stock1989new} are using it to estimate a monthly economic activity index for the U.S.. We include a measure of oil prices to capture the fact that energy-consuming states may behave differently to energy-producing states (e.g., see \cite{Carlino1999}).

In addition to these U.S.-level variables, we consider six predictors observed at the state level. Four of these indicators are those used by \cite{crone2005consistent} in the production of the Philadelphia Fed's state-level coincident indices. These are quarterly wages and salaries (including proprietors' income\footnote{In January 2020 the Federal Reserve Bank of Philadelphia considered an updated series for wages and salaries, that also included proprietors' income, when calculating its state coincident indices. We follow this recent practice.}), monthly nonfarm payroll employment, monthly average weekly hours worked, and the monthly unemployment rate. We then add in two additional indicators: quarterly real personal income and monthly initial claims.  It is important to consider personal income, given that while GDP is ostensibly an expenditure-side estimate, in practice the BEA relied on income-side data to measure state-level economic activity prior to 1997. And of course, even after 1997, we should expect personal income to correlate highly with state GDP. Initial claims are included as a timely indicator of movement in the labor market, a series consulted widely during the COVID-19 pandemic.

The exact definition, data source, transformation, and release schedule for each of these U.S. and state-level variables is given in the Data Appendix (Appendix \ref{DataAppendix}). We emphasize that due to differing publication lags these data have ragged-edges at the end of the sample. We accommodate these data features in our out-of-sample analysis by using the Kalman filter to interpolate endogenous variables missing at the end of the sample. Any missing exogenous state-level data are nowcast using recursively estimated AR(1) models.\footnote{There are, as detailed in the online data Appendix \ref{DataAppendix}, some instances of a ``ragged-head'' at the beginning of our sample. The most important are for the state-level unemployment rate and for state-level initial claims, both variables treated as exogenous in our MF-VAR. We fill-in these missing values at the beginning of our sample using the fitted values of an OLS autoregression also conditioning on the associated national (U.S.) variable.}

\subsection{In-Sample Analysis: Historical Estimates of State GDP}

We estimate our MF-VAR model on the latest (at the time of writing, June 2024) vintage data to produce monthly estimates of state GDP growth from the late 1960s through 2024m3. We remind the reader that we update these estimates in real time (each month) and they will be made available online as a resource to economists studying regional business cycles. To illustrate the utility of these data, below we show how they can be used to study regional business cycle dynamics and understand regional dependencies.

\subsubsection{Historical Estimates of Monthly GDP Across the U.S. States}

\begin{figure}
  \caption{Historical monthly estimates of state-level GDP growth, presented as year-on-year growth rates ($y^s_{a,t}$), for nine selected states}
  \centering
\includegraphics[scale=0.38]{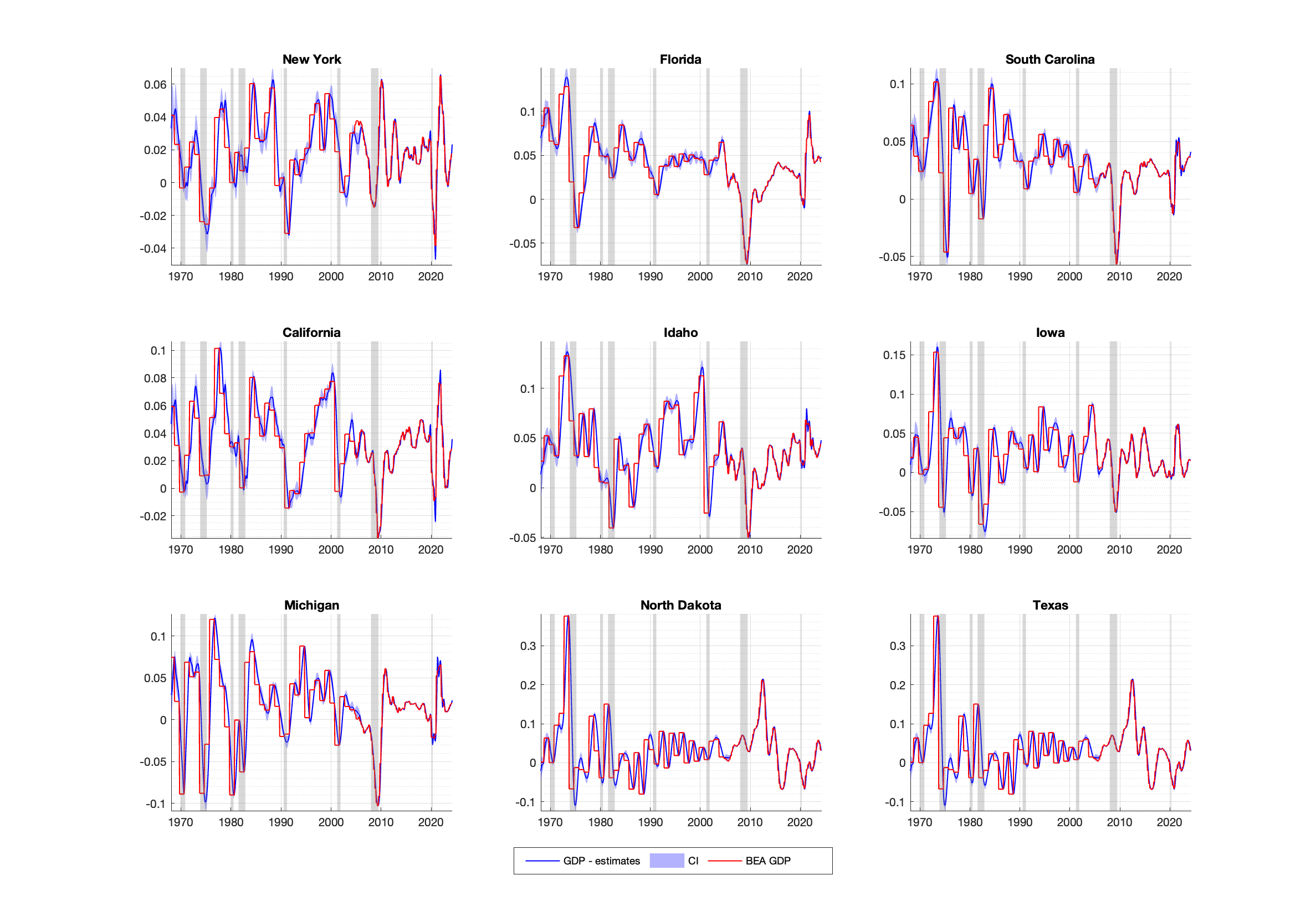}
    \label{fig:sreg}
    \footnotesize{Notes: NBER recession bands in gray. 68 percent credible intervals around the posterior median estimates shown in blue. Red lines denote BEA state-level GDP growth rate, available at the annual frequency prior to 2005, quarterly thereafter.}
\end{figure}

Figure \ref{fig:sreg} starts by illustrating the properties of the new monthly state-level GDP data for nine selected states representing nine Census Bureau divisions and regions of the U.S.. These are New York, representing the Northeast, Florida and South Carolina, representing the South Atlantic division, California representing the Pacific, Idaho representing the Mountain division, Iowa and Michigan
representing the Midwest, and North Dakota and Texas representing the oil-producing states of West North Central and West South Central. 

The posterior median of the state-level GDP growth estimates from our estimated MF-VAR model are plotted in Figure \ref{fig:sreg} as rolling monthly estimates of annual GDP growth, $y^s_{a,t}$. 68 percent credible intervals are also shown. Alongside our model-based estimates we show the BEA's annual and, from 2005, quarterly estimates of state GDP. As expected, given imposition of the temporal aggregation constraints, we see that once a year until 2005, and once a quarter thereafter, the model-based estimates align with the BEA estimates. We also, as expected given the more limited and lower frequency data available prior to 2005, see wider credible intervals around the model-based estimates until 2005. Figure \ref{fig:sreg} also evidences that while the state GDP growth estimates do co-move with the aggregate (U.S.) economy, there are heterogeneities across the states. 

\begin{figure}
    \caption{Correlation coefficients between each state GDP growth and U.S. growth (expressed as rolling monthly quarter-on-quarter growth rates) }
\centering
\includegraphics[scale=0.17]{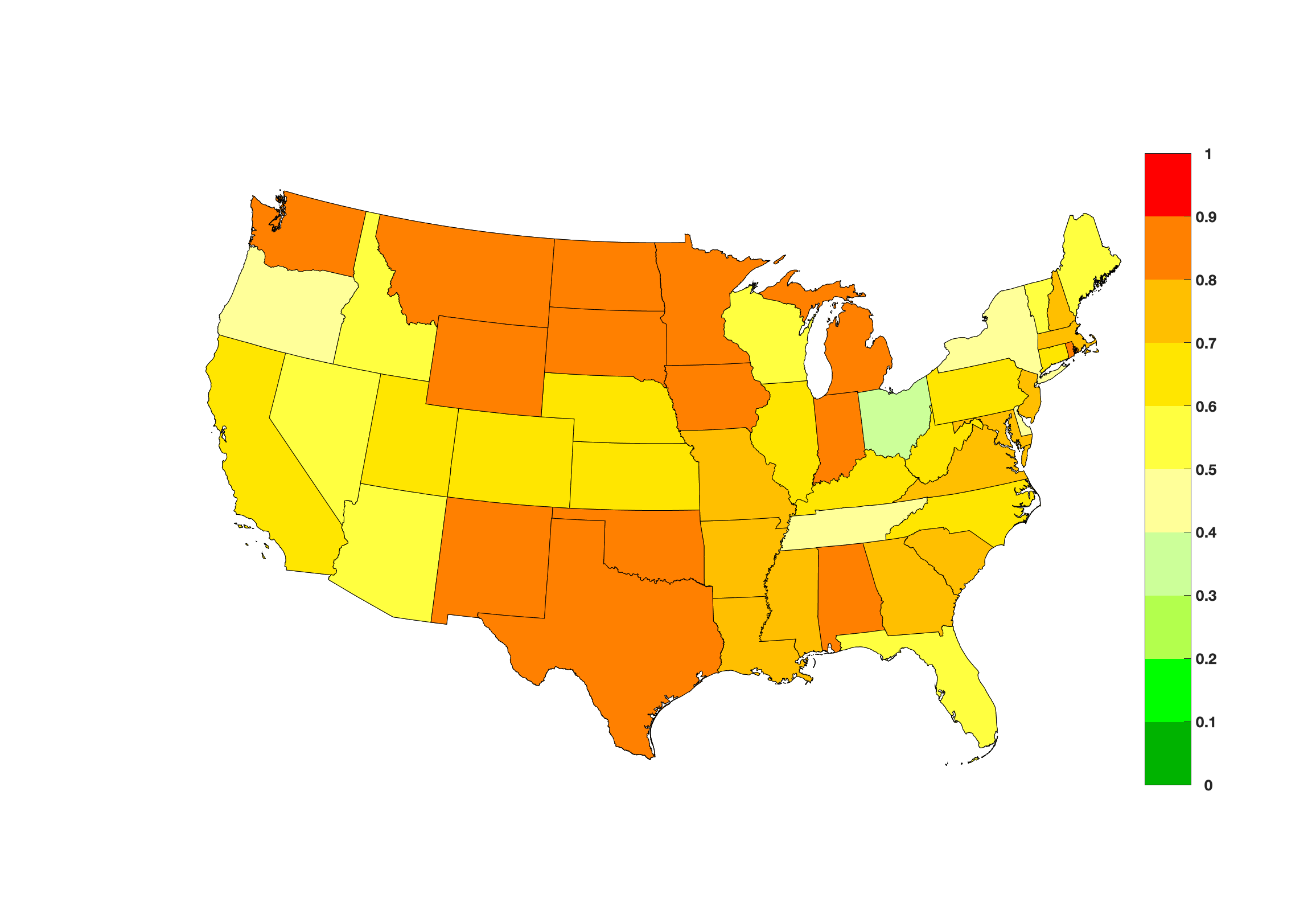}
    \label{fig:corr}
\footnotesize{Notes: Alaska, DC, and Hawaii, not plotted for space reasons, have correlation coefficients of 0.06, 0.53, and 0.57, respectively.}
\end{figure}

We further evidence state-level heterogeneity by, in Figure \ref{fig:corr}, reporting the correlation coefficient between each state's monthly GDP growth estimates (quarter-on-quarter) and the U.S.. This reveals that while GDP growth in Indiana (0.86), Illinois (0.88), and Pennsylvania (0.89) correlates highly with the U.S., states such as Alaska (0.06), North Dakota (0.32), and Wyoming (0.34) exhibit far weaker relationships, indicative of their different economic structures.

\subsubsection{State Business Cycle Dynamics}

\begin{figure}
\centering
\caption{Number of recessions in each state since 1964}
\hspace{-2.5cm} 
\includegraphics[scale=0.32]{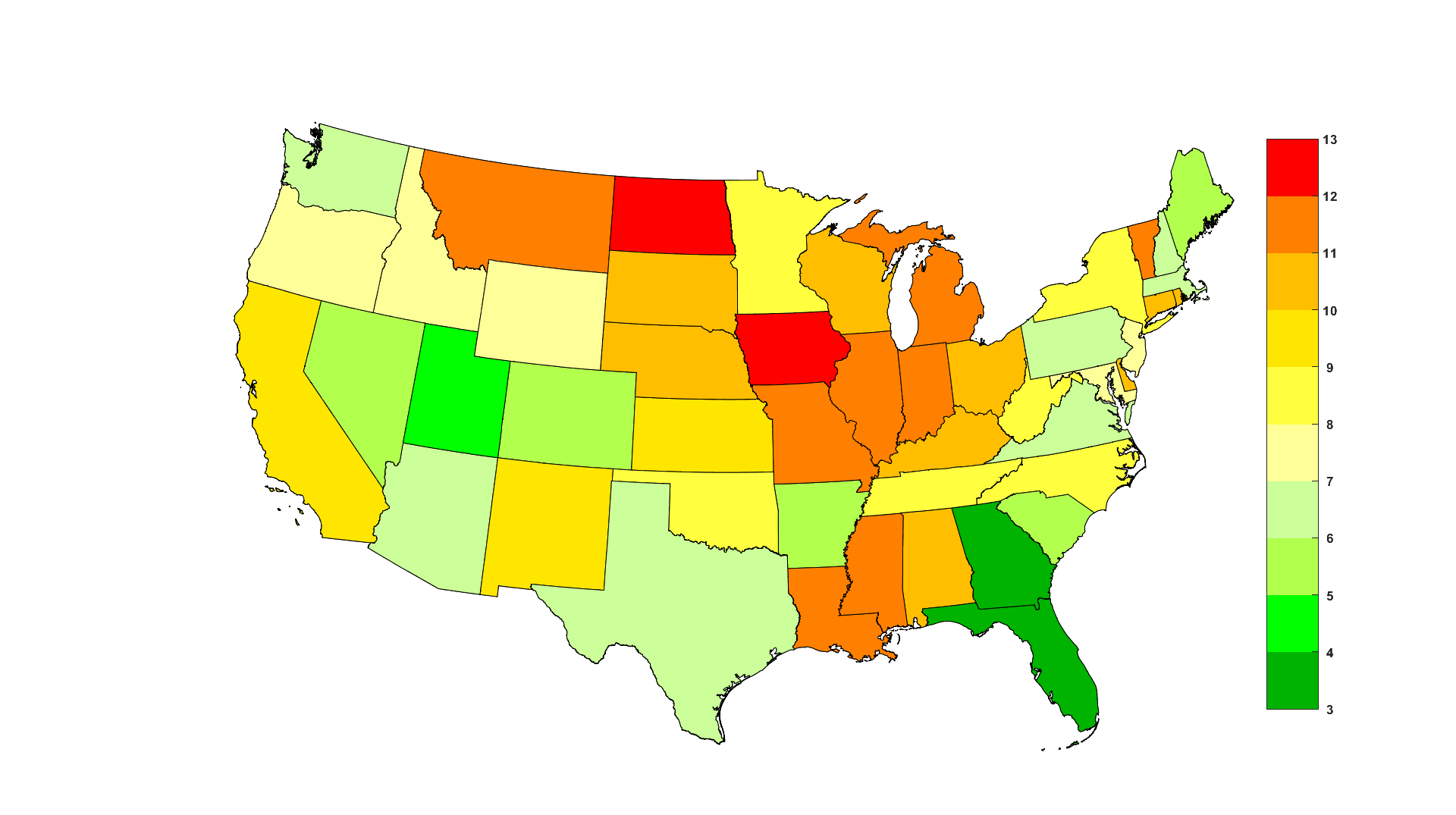}
\footnotesize{Notes: Recessions classified applying the \cite{HardingPagan} algorithm to the posterior median of the monthly state GDP estimates from the MF-VAR, (\ref{VAR}). Alaska, DC, and Hawaii, not plotted for space reasons, experienced 14, 10, and 10 recessions, respectively.}
    \label{fig:map}
\end{figure}

To draw out further common and contrasting features of these state business cycles, we apply the nonparametric business cycle dating algorithm of \cite{HardingPagan} to the median historical estimates of monthly state GDP growth (having transformed the data back into log-levels) to identify the turning points that separate state business cycle expansions from contractions. Figure \ref{fig:map} plots, by state, the number of recessions identified by this algorithm. This figure shows considerable variation as to the frequency of recessions, with Florida and Georgia experiencing the fewest (just three), and Iowa and North Dakota experiencing 13 recessions (and Alaska the most, at 14).

\subsubsection{Higher-Frequency Cross-State Dependencies}

The fact that our model generalizes previous work by jointly modelling the 51 states allows us to examine their dynamic connections. We use connectedness measures developed in \cite{diebold2014network} to investigate the dynamic connections between the U.S. states and U.S. macroeconomic variables. We do so at a higher frequency (monthly) than possible when using existing annual and, since 2005, quarterly state-level GDP data. \cite{diebold2014network} work off the generalized variance decomposition, following \cite{koop1996impulse} and \cite{pesaran1998generalized}, due to its invariance to the ordering of the variables in the VAR. We calculate the proportion of the $h$-step-ahead forecast error for GDP growth in state $n$ which is accounted for by the errors in the equation for variable $j$. Variable $j$ could be the equation for another state or for one of the macroeconomic variables. We denote the $nj$-th $h$-step-ahead proportion, $d_{nj}^h$. 

Following \cite{diebold2014network}, we define the total directional connectedness from other states to state $n$ at horizon $h$ as:
\begin{equation}
    \text{Connectedness from:} \sum_{n \neq j} d_{nj}^h.
\end{equation}

This is a measure of how information in other states impacts the forecast error variance of region $n$ (that is, the summation is over $j$). 

Then we define the total directional connectedness to other regions from region $j$ at horizon h as:
\begin{equation}
    \text{Connectedness to:} \sum_{j \neq n} d_{nj}^h.
\end{equation}

This is a measure of how information in state $j$ impacts the forecast error variances of other states (that is, the summation is over $n$). This is called a connectedness to measure.

 The connectedness from and to measures can be evaluated at each MCMC draw. The average of these produces the posterior mean which we use as the estimate.

\begin{sidewaysfigure}
\centering
  \caption{Full-sample \cite{diebold2014network} (DY) connectedness ``from'' and ``to'' estimates (variance decompositions, in percent). ``From'' estimates comprise effects from both the macroeconomic variables (in the MF-VAR) and other states.}
\hspace{-4.49cm}
\includegraphics[scale=.47]{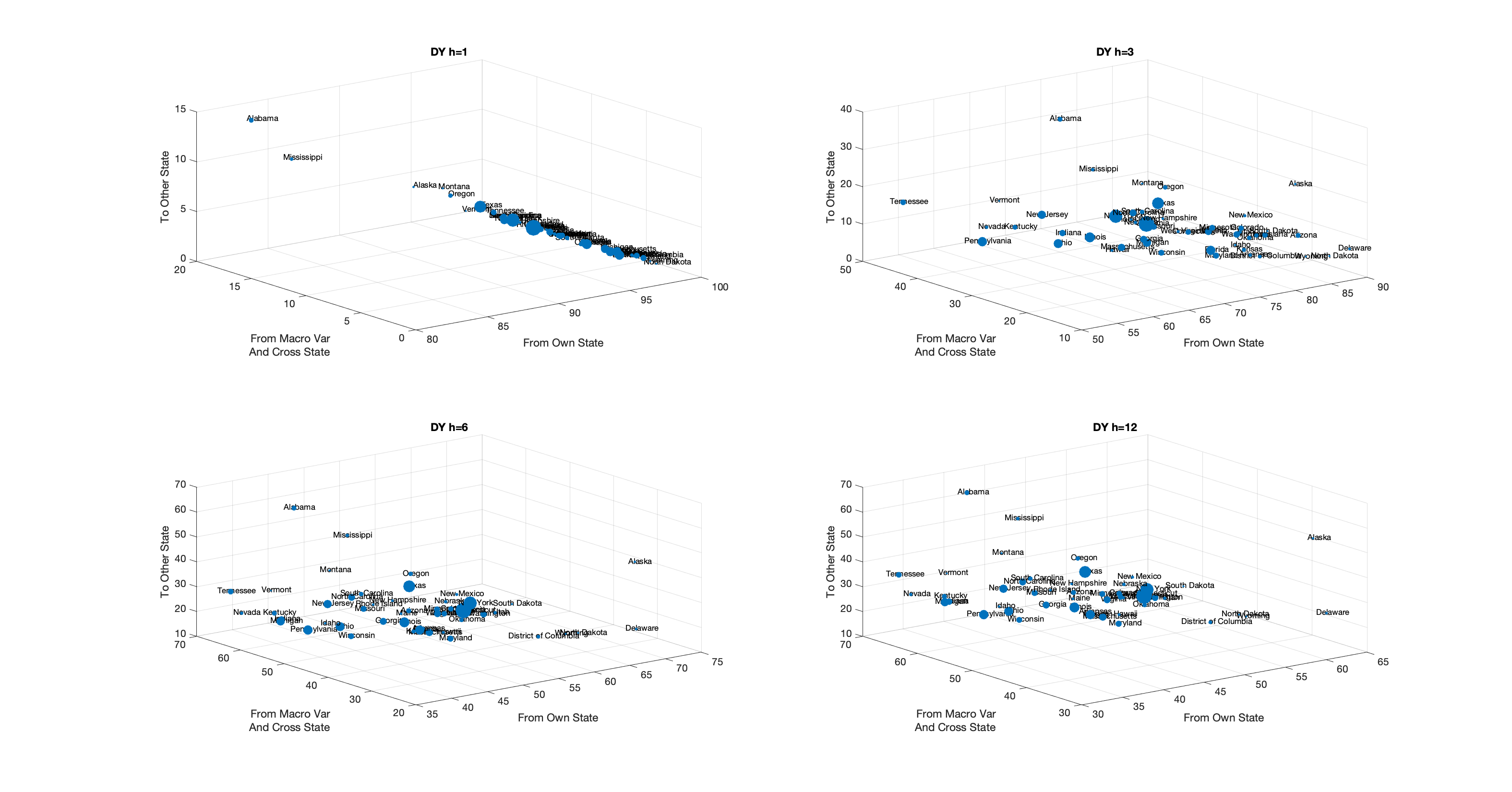}
    \label{fig:dyall}\\
    \footnotesize{Notes: Sample 1964m1 through 2024m3. Dots representing each state are sized to reflect the relative size of each state, as discussed in Footnote 9.}
\end{sidewaysfigure}

To illustrate how connectedness across states varies over time, in Figure \ref{fig:dyall} we plot four three-dimensional graphs for $h=1,3,6,12$ months. The first two dimensions decompose the ``connectedness from'' variance decompositions into their ``own state'' and macro plus cross-state contributions. The third dimension then plots ``connectedness to.''

As we move from the $h=1$ panel in Figure \ref{fig:dyall} through $h=12$, we see that the patterns of connectedness do change quite dramatically within the year, indeed within a quarter. At $h=1$, as we might expect, states tend to be most affected by idiosyncratic (state-specific) shocks. Alabama, Alaska, Montana, Mississippi, and Oregon are exceptions, showing greater sensitivity to macro shocks and to shocks from other states. Of these 5 states, Alabama and Mississippi stand out as they are the only two states (of all 50, plus DC) that at $h=1$ have noticeable, albeit quite small relative to the values at higher $h$, effects on the other states. But even just three months after the shock, we see far more connectedness, with the own-state effects for many states becoming much smaller. States such as Alaska and Delaware remain more affected by shocks specific to their state. In contrast, Pennsylvania and Tennessee, even after just three months, display strong sensitivity to macro- and cross-state spillovers, suggesting that they are more influenced by external economic conditions.

\subsection{Out-of-Sample Analysis: Timely Estimates and Nowcasts of State GDP Growth}

 In addition to producing historical monthly state GDP data, our MF-VAR model can be used out-of-sample (in real time) to produce more timely estimates and nowcasts of state GDP on an ongoing basis. These model-based real time estimates and nowcasts can be evaluated against the official state GDP data from the BEA once, at a greater lag, they are published. 

By way of establishing terminology, we produce: 1) three ``nowcasts'' of state GDP in quarter $\tau$ at the end of months 1, 2, and 3 of calendar quarter $\tau$, denoted m1, m2, and m3. We also produce (backward-looking) ``estimates'' of state GDP in quarter $\tau-1$ at the ends of months 1 and 2 of quarter $\tau$. These two estimates mimic and are to the same release delay as the BEA's publication of advance and first ``estimates'' for U.S. GDP growth. It is not necessary to make a third estimate, at m3, since by the end of the third month of a given quarter, state GDP for the previous calendar quarter is already known.\footnote{This is according to the current release schedule. Back in 2015 the BEA was releasing the quarterly estimates of state GDP 6 months after the end of the quarter. Over time, the BEA has sped up production of its state GDP estimates.} We should expect accuracy to improve as we move across these three nowcasts to the two estimates, since we are progressively conditioning on more and more information. ``Estimates'' should be more accurate than ``nowcasts.''

\begin{table}[t]
  \thispagestyle{empty} 
\caption{RMSE for quarterly GDP estimates and nowcasts, 2007Q1-2024Q1}
\label{RMSEtb1}
\resizebox{.93\columnwidth}{!}{%
    \begin{tabular}{@{}lccccc@{}}
\toprule
                       & m1 nowcast $\tau$ & m2 nowcast $\tau$ & m3 nowcast $\tau$ & m1 estimate $\tau-1$ & m2 estimate $\tau-1$ \\ \midrule
U.S.                    & 1.40              & 1.40              & 1.00              & --                 & --                 \\
Alabama                & 1.40              & 1.41              & 1.08              & 1.08                 & 1.07                 \\
Alaska                 & 2.33              & 2.41              & 2.19              & 2.17                 & 2.27                 \\
Arizona                & 2.23              & 2.23              & 1.94              & 1.92                 & 1.93                 \\
Arkansas               & 1.81              & 1.79              & 1.22              & 1.21                 & 1.19                 \\
California             & 2.18              & 2.18              & 1.77              & 1.80                 & 1.76                 \\
Colorado               & 1.57              & 1.57              & 1.30              & 1.26                 & 1.29                 \\
Connecticut            & 1.99              & 2.01              & 1.82              & 1.80                 & 1.80                 \\
Delaware               & 2.54              & 2.56              & 2.41              & 2.45                 & 2.49                 \\
District of Columbia & 1.06              & 1.05              & 1.01              & 1.01                 & 1.01                 \\
Florida                & 1.99              & 1.99              & 1.42              & 1.38                 & 1.39                 \\
Georgia                & 1.92              & 1.92              & 1.45              & 1.44                 & 1.44                 \\
Hawaii                 & 1.47              & 1.49              & 1.22              & 1.18                 & 1.21                 \\
Idaho                  & 1.51              & 1.52              & 1.39              & 1.41                 & 1.40                 \\
Illinois               & 1.35              & 1.35              & 1.04              & 1.03                 & 1.02                 \\
Indiana                & 1.62              & 1.60              & 1.33              & 1.30                 & 1.29                 \\
Iowa                   & 1.43              & 1.45              & 1.29              & 1.23                 & 1.22                 \\
Kansas                 & 1.62              & 1.61              & 1.40              & 1.37                 & 1.36                 \\
Kentucky               & 1.41              & 1.40              & 1.27              & 1.27                 & 1.27                 \\
Louisiana              & 1.73              & 1.73              & 1.42              & 1.42                 & 1.42                 \\
Maine                  & 1.29              & 1.29              & 1.10              & 1.09                 & 1.10                 \\
Maryland               & 1.61              & 1.61              & 1.35              & 1.34                 & 1.34                 \\
Massachusetts          & 1.48              & 1.47              & 1.24              & 1.24                 & 1.24                 \\
Michigan               & 1.74              & 1.70              & 1.47              & 1.34                 & 1.31                 \\
Minnesota              & 1.54              & 1.53              & 1.22              & 1.20                 & 1.21                 \\
Mississippi            & 1.50              & 1.50              & 1.22              & 1.23                 & 1.23                 \\
Missouri               & 1.46              & 1.43              & 1.18              & 1.16                 & 1.18                 \\
Montana                & 1.29              & 1.32              & 1.22              & 1.22                 & 1.22                 \\
Nebraska               & 1.54              & 1.57              & 1.83              & 1.78                 & 1.78                 \\
Nevada                 & 2.43              & 2.44              & 1.78              & 1.71                 & 1.72                 \\
New Hampshire         & 2.35              & 2.37              & 2.24              & 2.18                 & 2.18                 \\
New Jersey            & 1.51              & 1.50              & 1.25              & 1.24                 & 1.24                 \\
New Mexico            & 2.00              & 1.98              & 1.68              & 1.69                 & 1.67                 \\
New York              & 1.98              & 1.98              & 1.71              & 1.76                 & 1.73                 \\
North Carolina        & 1.82              & 1.81              & 1.32              & 1.34                 & 1.33                 \\
North Dakota          & 2.26              & 2.26              & 2.18              & 2.10                 & 2.09                 \\
Ohio                   & 1.25              & 1.25              & 1.09              & 1.08                 & 1.08                 \\
Oklahoma               & 1.55              & 1.53              & 1.57              & 1.57                 & 1.58                 \\
Oregon                 & 1.86              & 1.86              & 1.52              & 1.57                 & 1.56                 \\
Pennsylvania           & 1.31              & 1.31              & 1.11              & 1.11                 & 1.12                 \\
Rhode Island          & 1.63              & 1.65              & 1.40              & 1.38                 & 1.37                 \\
South Carolina        & 1.60              & 1.60              & 1.28              & 1.26                 & 1.27                 \\
South Dakota          & 2.18              & 2.19              & 2.55              & 2.49                 & 2.49                 \\
Tennessee              & 1.52              & 1.52              & 1.19              & 1.22                 & 1.22                 \\
Texas                  & 1.96              & 1.95              & 1.59              & 1.53                 & 1.51                 \\
Utah                   & 1.67              & 1.67              & 1.44              & 1.42                 & 1.43                 \\
Vermont                & 1.57              & 1.61              & 1.33              & 1.32                 & 1.35                 \\
Virginia               & 1.71              & 1.72              & 1.28              & 1.28                 & 1.28                 \\
Washington             & 1.65              & 1.66              & 1.47              & 1.44                 & 1.43                 \\
West Virginia         & 1.32              & 1.31              & 1.34              & 1.36                 & 1.34                 \\
Wisconsin              & 1.42              & 1.42              & 1.10              & 1.07                 & 1.07                 \\
Wyoming                & 2.03              & 2.02              & 2.04              & 2.00                 & 2.02                 \\ \bottomrule
AVG                    & 1.71              & 1.71              & 1.48              & 1.46                 & 1.46                 \\ \bottomrule
\end{tabular}
}\\
\footnotesize{Notes: RMSE statistics x 100 for nowcasts for quarter $\tau$ made at the end of month 1, 2, or 3 of quarter $\tau$ and estimates for quarter $\tau-1$ made at the end of month 1 and 2 of quarter $\tau$. AVG denotes the equal-weighted average across all states. 2020Q2-Q4 dropped when computing RMSE to avoid contamination due to COVID-19 outliers.}
\end{table}

\begin{table}[]
\caption{Average CRPS for quarterly GDP estimates and nowcasts, 2007Q1-2024Q1}
\label{CRPStb1}
\resizebox{.93\columnwidth}{!}{%
\begin{tabular}{@{}lccccc@{}}
\toprule
                       & m1 nowcast $\tau$ & m2 nowcast $\tau$ & m3 nowcast $\tau$ & m1 estimate $\tau-1$ & m2 estimate $\tau-1$ \\ \midrule
U.S.                    & 0.99              & 1.00              & 0.58              & --                 & --                \\
Alabama                & 0.89              & 0.90              & 0.62              & 0.62                 & 0.63                 \\
Alaska                 & 1.79              & 1.82              & 1.48              & 1.43                 & 1.52                 \\
Arizona                & 1.53              & 1.54              & 1.15              & 1.14                 & 1.16                 \\
Arkansas               & 1.14              & 1.13              & 0.70              & 0.70                 & 0.69                 \\
California             & 1.58              & 1.58              & 1.14              & 1.15                 & 1.13                 \\
Colorado               & 1.03              & 1.03              & 0.75              & 0.72                 & 0.74                 \\
Connecticut            & 1.26              & 1.28              & 1.03              & 1.00                 & 1.02                 \\
Delaware               & 1.63              & 1.64              & 1.43              & 1.45                 & 1.50                 \\
District of Columbia & 0.65              & 0.65              & 0.58              & 0.58                 & 0.59                 \\
Florida                & 1.36              & 1.37              & 0.86              & 0.84                 & 0.85                 \\
Georgia                & 1.29              & 1.30              & 0.83              & 0.83                 & 0.83                 \\
Hawaii                 & 0.90              & 0.91              & 0.70              & 0.67                 & 0.69                 \\
Idaho                  & 1.01              & 1.02              & 0.81              & 0.81                 & 0.82                 \\
Illinois               & 0.84              & 0.84              & 0.58              & 0.58                 & 0.58                 \\
Indiana                & 1.03              & 1.02              & 0.77              & 0.77                 & 0.76                 \\
Iowa                   & 0.90              & 0.92              & 0.77              & 0.74                 & 0.75                 \\
Kansas                 & 0.95              & 0.94              & 0.77              & 0.75                 & 0.76                 \\
Kentucky               & 0.89              & 0.89              & 0.73              & 0.73                 & 0.74                 \\
Louisiana              & 1.04              & 1.04              & 0.78              & 0.79                 & 0.80                 \\
Maine                  & 0.79              & 0.78              & 0.64              & 0.64                 & 0.65                 \\
Maryland               & 1.04              & 1.05              & 0.78              & 0.79                 & 0.78                 \\
Massachusetts          & 0.99              & 0.98              & 0.74              & 0.74                 & 0.76                 \\
Michigan               & 1.10              & 1.08              & 0.86              & 0.80                 & 0.80                 \\
Minnesota              & 0.94              & 0.93              & 0.70              & 0.69                 & 0.70                 \\
Mississippi            & 0.97              & 0.97              & 0.73              & 0.74                 & 0.75                 \\
Missouri               & 0.88              & 0.87              & 0.64              & 0.65                 & 0.66                 \\
Montana                & 0.85              & 0.86              & 0.74              & 0.75                 & 0.76                 \\
Nebraska               & 0.98              & 1.00              & 1.08              & 1.04                 & 1.05                 \\
Nevada                 & 1.67              & 1.69              & 1.13              & 1.08                 & 1.10                 \\
New Hampshire         & 1.55              & 1.58              & 1.43              & 1.39                 & 1.40                 \\
New Jersey            & 0.98              & 0.97              & 0.72              & 0.72                 & 0.72                 \\
New Mexico            & 1.34              & 1.33              & 0.99              & 1.00                 & 1.01                 \\
New York              & 1.27              & 1.27              & 1.02              & 1.04                 & 1.03                 \\
North Carolina        & 1.20              & 1.20              & 0.77              & 0.79                 & 0.79                 \\
North Dakota          & 1.63              & 1.62              & 1.37              & 1.35                 & 1.37                 \\
Ohio                   & 0.78              & 0.78              & 0.60              & 0.60                 & 0.60                 \\
Oklahoma               & 0.95              & 0.95              & 0.91              & 0.90                 & 0.91                 \\
Oregon                 & 1.24              & 1.24              & 0.90              & 0.94                 & 0.94                 \\
Pennsylvania           & 0.82              & 0.82              & 0.63              & 0.63                 & 0.64                 \\
Rhode Island          & 1.01              & 1.02              & 0.83              & 0.82                 & 0.83                 \\
South Carolina        & 1.03              & 1.04              & 0.76              & 0.74                 & 0.75                 \\
South Dakota          & 1.42              & 1.43              & 1.57              & 1.54                 & 1.56                 \\
Tennessee              & 0.99              & 0.99              & 0.68              & 0.69                 & 0.71                 \\
Texas                  & 1.38              & 1.37              & 1.01              & 0.97                 & 0.96                 \\
Utah                   & 1.11              & 1.12              & 0.87              & 0.86                 & 0.87                 \\
Vermont                & 1.06              & 1.08              & 0.80              & 0.80                 & 0.83                 \\
Virginia               & 1.18              & 1.19              & 0.78              & 0.78                 & 0.79                 \\
Washington             & 1.03              & 1.03              & 0.84              & 0.83                 & 0.83                 \\
West Virginia         & 0.81              & 0.81              & 0.78              & 0.80                 & 0.80                 \\
Wisconsin              & 0.89              & 0.89              & 0.62              & 0.61                 & 0.62                 \\
Wyoming                & 1.25              & 1.25              & 1.23              & 1.21                 & 1.24      \\ \bottomrule           
AVG                    & 1.11              & 1.12              & 0.88              & 0.87                 & 0.88       \\ \bottomrule
\end{tabular}
}
\\
\footnotesize{Notes: CRPS statistics x 100. AVG denotes the equal-weighted average across all states. 2020Q2-Q4 dropped when computing average CRPS to avoid contamination due to COVID-19 outliers.}

\end{table}

\begin{table}[]
\caption{RMSE ratio relative to benchmark state-specific MF-VAR}
\label{tab3}
\resizebox{.93\columnwidth}{!}{%
\begin{tabular}{@{}lccccc@{}}
\toprule
                       & m1 nowcast $\tau$ & m2 nowcast $\tau$ & m3 nowcast $\tau$ & m1 estimate $\tau-1$ & m2 estimate $\tau-1$ \\ \midrule
U.S.                    & 0.88              & 0.88              & 0.80              &  --                    &   --                   \\
Alabama                & 0.94              & 0.94              & 0.89              & 0.93                 & 0.92                 \\
Alaska                 & 0.83              & 0.84              & 0.81              & 0.81                 & 0.83                 \\
Arizona                & 1.09              & 1.09              & 0.86              & 0.87                 & 0.87                 \\
Arkansas               & 1.02              & 1.02              & 0.95              & 0.99                 & 0.93                 \\
California             & 1.49              & 1.48              & 1.24              & 1.27                 & 1.20                 \\
Colorado               & 1.08              & 1.07              & 1.01              & 1.01                 & 1.00                 \\
Connecticut            & 1.05              & 1.04              & 0.96              & 0.99                 & 0.94                 \\
Delaware               & 1.06              & 1.05              & 0.98              & 1.02                 & 0.96                 \\
District of Columbia & 1.01              & 1.02              & 0.86              & 0.86                 & 0.84                 \\
Florida                & 1.22              & 1.21              & 1.13              & 1.15                 & 1.09                 \\
Georgia                & 1.11              & 1.10              & 1.01              & 1.04                 & 1.00                 \\
Hawaii                 & 0.91              & 0.91              & 1.00              & 1.01                 & 0.95                 \\
Idaho                  & 1.03              & 1.03              & 0.87              & 0.89                 & 0.88                 \\
Illinois               & 1.00              & 1.00              & 0.86              & 0.89                 & 0.83                 \\
Indiana                & 0.95              & 0.93              & 0.82              & 0.83                 & 0.80                 \\
Iowa                   & 0.94              & 0.93              & 0.81              & 0.82                 & 0.77                 \\
Kansas                 & 0.98              & 0.98              & 0.90              & 0.92                 & 0.91                 \\
Kentucky               & 0.91              & 0.90              & 0.84              & 0.86                 & 0.84                 \\
Louisiana              & 0.92              & 0.91              & 1.00              & 1.01                 & 0.98                 \\
Maine                  & 0.91              & 0.89              & 0.80              & 0.82                 & 0.80                 \\
Maryland               & 1.01              & 1.01              & 0.97              & 0.98                 & 0.91                 \\
Massachusetts          & 1.05              & 1.05              & 0.99              & 1.00                 & 0.96                 \\
Michigan               & 0.92              & 0.90              & 0.93              & 0.89                 & 0.86                 \\
Minnesota              & 1.00              & 1.01              & 0.78              & 0.79                 & 0.75                 \\
Mississippi            & 0.97              & 0.95              & 0.82              & 0.86                 & 0.82                 \\
Missouri               & 0.97              & 0.98              & 0.86              & 0.85                 & 0.83                 \\
Montana                & 0.86              & 0.85              & 0.80              & 0.81                 & 0.77                 \\
Nebraska               & 0.92              & 0.93              & 0.84              & 0.85                 & 0.84                 \\
Nevada                 & 1.18              & 1.17              & 1.13              & 1.15                 & 1.11                 \\
New Hampshire         & 1.05              & 1.05              & 0.88              & 0.88                 & 0.88                 \\
New Jersey            & 1.07              & 1.06              & 0.96              & 0.96                 & 0.92                 \\
New Mexico            & 1.06              & 1.06              & 0.98              & 0.99                 & 0.98                 \\
New York              & 1.16              & 1.17              & 0.95              & 1.01                 & 0.99                 \\
North Carolina        & 1.03              & 1.02              & 0.89              & 0.91                 & 0.86                 \\
North Dakota          & 0.87              & 0.86              & 0.85              & 0.84                 & 0.83                 \\
Ohio                   & 0.98              & 0.96              & 0.92              & 0.97                 & 0.88                 \\
Oklahoma               & 1.04              & 1.04              & 0.93              & 0.94                 & 0.93                 \\
Oregon                 & 1.03              & 1.03              & 0.89              & 0.92                 & 0.87                 \\
Pennsylvania           & 0.98              & 0.98              & 0.94              & 0.95                 & 0.93                 \\
Rhode Island          & 0.97              & 0.98              & 0.86              & 0.89                 & 0.86                 \\
South Carolina        & 0.98              & 0.98              & 0.92              & 0.96                 & 0.92                 \\
South Dakota          & 0.95              & 0.95              & 0.80              & 0.80                 & 0.80                 \\
Tennessee              & 0.96              & 0.95              & 0.86              & 0.89                 & 0.88                 \\
Texas                  & 1.16              & 1.14              & 1.15              & 1.13                 & 1.11                 \\
Utah                   & 1.02              & 1.02              & 0.99              & 1.01                 & 1.00                 \\
Vermont                & 0.97              & 0.97              & 0.88              & 0.89                 & 0.87                 \\
Virginia               & 1.01              & 1.02              & 1.00              & 1.03                 & 1.00                 \\
Washington             & 1.04              & 1.05              & 1.01              & 1.01                 & 1.00                 \\
West Virginia         & 0.94              & 0.94              & 0.89              & 0.90                 & 0.88                 \\
Wisconsin              & 0.95              & 0.94              & 0.84              & 0.86                 & 0.83                 \\
Wyoming                & 0.98              & 0.97              & 0.92              & 0.92                 & 0.92                 \\ \bottomrule
AVG                    & 1.01              & 1.00              & 0.91              & 0.93                 & 0.90             \\ \bottomrule
\end{tabular}
}
\footnotesize{Notes: Ratios less than one indicate higher forecast accuracy for the cross-state MF-VAR model, (\ref{VAR}), than the state-specific MF-VAR model. 2020Q2-Q4 dropped to avoid contamination due to COVID-19 outliers.}
\end{table}

\begin{table}[]
\caption{CRPS ratio relative to benchmark state-specific MF-VAR}
\label{tab4}
\resizebox{.93\columnwidth}{!}{%
\begin{tabular}{@{}lccccc@{}}
\toprule
                       & m1 nowcast $\tau$ & m2 nowcast $\tau$ & m3 nowcast $\tau$ & m1 estimate $\tau-1$ & m2 estimate $\tau-1$ \\ \midrule
U.S.                    & 0.86              & 0.86              & 0.72              &   --                   &   --                   \\
Alabama                & 0.79              & 0.78              & 0.78              & 0.84                 & 0.78                 \\
Alaska                 & 0.87              & 0.88              & 0.81              & 0.80                 & 0.76                 \\
Arizona                & 1.00              & 1.00              & 0.82              & 0.86                 & 0.81                 \\
Arkansas               & 0.86              & 0.85              & 0.82              & 0.88                 & 0.81                 \\
California             & 1.53              & 1.51              & 1.28              & 1.34                 & 1.25                 \\
Colorado               & 0.94              & 0.93              & 0.93              & 0.94                 & 0.89                 \\
Connecticut            & 0.95              & 0.94              & 0.88              & 0.92                 & 0.87                 \\
Delaware               & 1.02              & 1.01              & 0.95              & 1.00                 & 0.94                 \\
District of Columbia & 0.85              & 0.85              & 0.81              & 0.82                 & 0.79                 \\
Florida                & 1.13              & 1.13              & 1.02              & 1.06                 & 0.97                 \\
Georgia                & 0.99              & 0.98              & 0.90              & 0.95                 & 0.89                 \\
Hawaii                 & 0.87              & 0.88              & 0.89              & 0.90                 & 0.86                 \\
Idaho                  & 0.88              & 0.88              & 0.76              & 0.80                 & 0.76                 \\
Illinois               & 0.86              & 0.85              & 0.76              & 0.79                 & 0.75                 \\
Indiana                & 0.82              & 0.81              & 0.72              & 0.76                 & 0.67                 \\
Iowa                   & 0.77              & 0.78              & 0.71              & 0.74                 & 0.69                 \\
Kansas                 & 0.84              & 0.83              & 0.82              & 0.85                 & 0.79                 \\
Kentucky               & 0.75              & 0.74              & 0.74              & 0.77                 & 0.72                 \\
Louisiana              & 0.83              & 0.82              & 0.87              & 0.90                 & 0.86                 \\
Maine                  & 0.76              & 0.74              & 0.72              & 0.74                 & 0.70                 \\
Maryland               & 0.88              & 0.89              & 0.88              & 0.91                 & 0.86                 \\
Massachusetts          & 0.95              & 0.95              & 0.92              & 0.95                 & 0.89                 \\
Michigan               & 0.90              & 0.89              & 0.86              & 0.85                 & 0.75                 \\
Minnesota              & 0.81              & 0.83              & 0.70              & 0.72                 & 0.68                 \\
Mississippi            & 0.81              & 0.80              & 0.75              & 0.80                 & 0.73                 \\
Missouri               & 0.81              & 0.82              & 0.76              & 0.78                 & 0.74                 \\
Montana                & 0.72              & 0.72              & 0.71              & 0.74                 & 0.70                 \\
Nebraska               & 0.79              & 0.80              & 0.74              & 0.76                 & 0.74                 \\
Nevada                 & 1.13              & 1.13              & 1.06              & 1.08                 & 1.03                 \\
New Hampshire         & 1.00              & 1.00              & 0.85              & 0.86                 & 0.83                 \\
New Jersey            & 0.95              & 0.94              & 0.86              & 0.87                 & 0.82                 \\
New Mexico            & 0.93              & 0.93              & 0.91              & 0.94                 & 0.89                 \\
New York              & 1.13              & 1.14              & 0.93              & 0.99                 & 0.95                 \\
North Carolina        & 0.89              & 0.88              & 0.79              & 0.83                 & 0.77                 \\
North Dakota          & 0.80              & 0.80              & 0.79              & 0.79                 & 0.77                 \\
Ohio                   & 0.84              & 0.82              & 0.80              & 0.85                 & 0.74                 \\
Oklahoma               & 0.96              & 0.96              & 0.87              & 0.89                 & 0.85                 \\
Oregon                 & 0.92              & 0.92              & 0.82              & 0.86                 & 0.80                 \\
Pennsylvania           & 0.85              & 0.85              & 0.83              & 0.86                 & 0.82                 \\
Rhode Island          & 0.87              & 0.88              & 0.77              & 0.81                 & 0.79                 \\
South Carolina        & 0.84              & 0.84              & 0.81              & 0.85                 & 0.81                 \\
South Dakota          & 0.85              & 0.85              & 0.73              & 0.75                 & 0.72                 \\
Tennessee              & 0.83              & 0.82              & 0.75              & 0.80                 & 0.74                 \\
Texas                  & 1.05              & 1.04              & 1.07              & 1.08                 & 1.02                 \\
Utah                   & 0.88              & 0.88              & 0.88              & 0.92                 & 0.88                 \\
Vermont                & 0.86              & 0.86              & 0.80              & 0.82                 & 0.78                 \\
Virginia               & 0.89              & 0.89              & 0.88              & 0.92                 & 0.88                 \\
Washington             & 0.91              & 0.91              & 0.90              & 0.92                 & 0.89                 \\
WestVirginia         & 0.80              & 0.79              & 0.81              & 0.83                 & 0.80                 \\
Wisconsin              & 0.79              & 0.78              & 0.73              & 0.76                 & 0.72                 \\
Wyoming                & 0.86              & 0.85              & 0.87              & 0.88                 & 0.81                 \\ \bottomrule
AVG                    & 0.90              & 0.90              & 0.84              & 0.86                 & 0.81   \\ \bottomrule
\end{tabular}
}
\footnotesize{Notes: Ratios less than one indicate higher forecast accuracy for the cross-state MF-VAR model, (\ref{VAR}), than the state-specific MF-VAR model. 2020Q2-Q4 dropped to avoid contamination due to COVID-19 outliers.}
\end{table}

Specifically, our timing convention is that, using the most recently available real time data vintage, we recursively rerun the MF-VAR model each month upon receipt of the latest U.S. GDP data from the BEA along with revised and new data for all other indicators.\footnote{In the spirit of \cite{SchorfheideSong2024}, when recursively estimating the model, we do not update the parameters of the MF-VAR for 9 months from March 2020 to ensure that the nowcasts, and underlying parameters in the MF-VAR, are not contaminated by the extreme COVID-19 observations. We also drop the latter three quarters of 2020 when computing our evaluation metrics. } The BEA publish their quarterly estimates and revisions for U.S. GDP towards the end of each month in quarter $\tau$, typically in the last week of the month. We also condition on the latest values of the other indicators in our model, known at this point (last week) in the month. Most of these relate to the previous month.\footnote{Table \ref{t3} in the online Appendix provides a typical release calendar for the U.S. and state variables used in our MF-VAR model.} This means that we can use our MF-VAR model to produce state GDP estimates at least two months ahead of the BEA. The nowcasts made at the end of m1 have an even greater, five month timing gain over the BEA.

We evaluate the accuracy of the MF-VAR's point and density estimates and nowcasts against the subsequent outturns from the BEA using the root mean square forecast error (RMSFE) and the continuous ranked probability score (CRPS), respectively. Lower values of each of these metrics indicate improved accuracy.\footnote{In online Appendix \ref{EmpiricsAppendix}, we also report results evaluating the densities using the logarithmic score. This yields similar conclusions to the CRPS.} As the BEA have produced quarterly state-level GDP estimates (dating back to 2005), this means that we can evaluate our monthly estimates four times a year, after aggregating our monthly estimates of state GDP growth to rolling (calendar) quarterly estimates of state GDP growth via Equation (\ref{q_to_m_states}). There is always a question about which release of state GDP to use as the ``outturn.'' We focus on the latest (June 2024) vintage estimates from the BEA. Our evaluation period runs from 2007Q1, since this is when, as detailed in the online data Appendix \ref{DataAppendix}, the real time data vintages date back to, through 2024Q1.

We also compare the accuracy of the density nowcasts and estimates from our MF-VAR model, (\ref{VAR}), against those from a benchmark model. The benchmark model is a restricted case of our MF-VAR model that neither models cross-state linkages nor imposes the cross-sectional aggregation constraint, (\ref{cross_sec}). Specifically, the benchmark model is a state-level MF-VAR with monthly employment, quarterly U.S. GDP and state GDP as endogenous variables, and the six state-level exogenous variables.

Tables \ref{RMSEtb1} and \ref{CRPStb1} show the RMSE and average CRPS statistics for our nowcast/estimate exercise. The top row of each table evaluates the accuracy of the nowcasts for U.S. GDP. Model-based ``estimates'' of U.S. GDP are not required, since the advance estimate of quarterly U.S. GDP is published at the end of the following month. Looking across the rows of these two tables, we see considerable cross-state variation in accuracy. Looking across the columns, as hoped, we see that accuracy tends to improve as information accumulates -- as we move to the right in each table. These gains are summarized in the final row of each table when reporting the RMSE and CRPS statistics averaged across states. These averaged statistics support our main finding that accuracy clearly improves in an absolute sense as more within-quarter and past-quarter information accumulates. They also indicate that the biggest jump in forecast accuracy, both for the point and density nowcasts and estimates, happens when information on the third month in the quarter (see the m3 nowcast) becomes available. Interestingly, this jump in accuracy occurs before the BEA publish their own advance estimate of quarterly U.S. GDP. That is, we find that there is little gain to waiting an extra month and using the advance estimate of state GDP rather than a model-based nowcast, as long as this nowcast conditions on two months of within-quarter information. This is understood when we observe that the accuracy of the nowcasts for U.S. GDP growth also improves dramatically at m3. This means, in effect, that when computing nowcasts of state GDP at m3, the model is conditioning, via the cross-sectional aggregation constraint, on \textit{good} estimates of U.S. GDP.

Tables \ref{tab3} and \ref{tab4} confirm that from the third month of the current quarter our (cross-state) MF-VAR model, (\ref{VAR}), consistently produces more accurate point and density nowcasts and estimates than the benchmark state-specific model. The RMSE and CRPS ratios in Tables \ref{tab3} and \ref{tab4} are consistently below unity, both at the state level and when averaged across states. This demonstrates the value-added of producing nowcasts and estimates of state GDP both conditioning on other states' GDP and imposing the cross-sectional aggregation constraint, (\ref{cross_sec}), that forces the estimates of state GDP to be consistent with those of U.S. GDP.

\section{Conclusion}

This paper develops a ``big data'' MF-VAR model that is used to produce both historical monthly estimates and more timely nowcasts of state GDP. The estimates and nowcasts are both cross-sectionally and temporally consistent with official lower-frequency, and less timely, state and U.S. GDP data from the BEA. Our model exploits monthly, quarterly, and annual data, both for state GDP, U.S. GDP, and state and U.S.-level indicator variables. It allows for dynamic interactions between all of these variables. Importantly, consistency between higher-frequency model-based and lower-frequency BEA data is imposed via temporal and cross-sectional constraints embedded within the MF-VAR.

Estimating a model jointly across states, rather than, as in previous research, estimating a state-specific mixed-frequency model, is challenging due to the high dimension of the model and the complicated data frequency mismatch that ensues. That is, the MF-VAR must have  $51$ equations for each of the $50$ states, plus DC, plus additional equations to model the macro variables at the U.S. level. Moreover, the state-level data change frequency in the sample,  from annual to quarterly. In this paper, we have shown how Bayesian methods involving the horseshoe prior are able to ensure shrinkage in this otherwise greatly over-parameterized model. We have overcome the computational challenge of undertaking MCMC in such high dimensions by developing a novel algorithm which handles the three-way frequency mismatch by combining two much simpler algorithms, each involving two-way frequency mismatches. 

In our empirical work, we find superior real time nowcast performance relative to a  benchmark which is an alternative mixed frequency VAR but does not allow for cross-state linkages. We find that the cross-state model produces accurate nowcasts of quarterly state GDP as soon as two months of within-quarter data are known. These nowcasts are available four month ahead of the BEA's first estimates for state GDP, providing economists with a considerably faster impression of state-level economic activity. We also illustrate how the historical monthly estimates of state 
 GDP that our model produces back to the 1960s can be used to gain understanding of state business cycle dynamics and their connectedness. The historical monthly state GDP estimates and updated nowcasts from our model  will be maintained and made available to others on our websites.

\clearpage
\newpage
\bibliography{KMMR}
\newpage
\clearpage

\appendix

\setcounter{equation}{1}
 \setcounter{page}{1}
\counterwithin{table}{section}
 \counterwithin{figure}{section}
 
\noindent{\LARGE{\textbf{Online Appendix}}}
\\

 \renewcommand{\thepage}{A\arabic{page}}
 \setcounter{page}{1}

\renewcommand{\thesection}{A.\arabic{section}}

Appendix \ref{DataAppendix} discusses  construction of the real time monthly, quarterly, and annual state and U.S. data set. Appendix \ref{TechAppendix} details implementation and computation details for the Bayesian MF-VAR model. Appendix \ref{EmpiricsAppendix} provides supplementary empirical results referred to in the main paper.

\section{Data Appendix}\label{DataAppendix}

Our mixed-frequency dataset consists of quarterly U.S. real GDP, annual and quarterly state-level real GDP, plus 13 U.S. macroeconomic indicators and 6 state-level variables available at the monthly or quarterly frequency.

Table \ref{Tab:Data} lists these 22 variables, indicating for each the data source, the frequency of the observed data, the data transformation used when modeling in the MF-VAR model, the historical sample period available, and the availability of data vintages used in the out-of-sample analysis. The sample used to estimate our MF-VAR models begins, at the earliest, in 1964 (in annual growth rates, 1963 in levels) because 1963 is the first year of data available for (nominal) state GDP. The state-level indicators of state GDP are often available over a shorter sample than the corresponding U.S. data.

Here we provide some additional details about the data used.

We follow the recommendations of \cite{mccracken2016fred} in selecting what data transformation to take. In Table \ref{Tab:Data}, we employ data transformations similar to those in \cite{mccracken2016fred} for nearly all listed series. Specifically, for real personal income, the IP Index, initial claims, all employees payroll: total nonfarm, government expenditure, and personal consumption expenditure, we apply the first log difference. For the civilian unemployment Rate, the effective federal funds Rate, and the 10-year Treasury rate, we use the first difference. The series for CPI: All items, crude oil, spliced WTI and Cushing, and fixed investment, are transformed using the second log difference. Notably, while \cite{mccracken2016fred} use levels data for average weekly hours: manufacturing, at both the national and state levels, we apply the first difference. Additionally, we use the second log difference for fixed investment and the first log difference for wages and salaries. A key difference from \cite{mccracken2016fred} is our treatment of the civilian unemployment rate, for which we use the first difference to smooth out the COVID-19 spike that caused unstable draws in our MCMC algorithm. 

We seasonally adjust any unadjusted series using the X-13 ARIMA-SEATS seasonal adjustment program from the Census Bureau. We use the default options for X11. When constructing the real-time data vintages seasonal adjustment is undertaken recursively, to each data vintage in turn, mimicking real-time application. 

Of the variables listed in Table \ref{Tab:Data}, the S\&P 500, the effective federal funds rate, the 10 year T-bill rate, and initial claims are in fact available at a higher frequency than monthly, the highest data frequency we exploit in our MF-VAR models. But to reflect the fact that these data are available at a higher frequency than monthly, when using these variables in the out-of-sample analysis, we do not take a monthly average for the most recent month. Instead, we take an average over the first 3 weeks of the latest month and compare to the previous (full) month's data. We consider data for only the first 3 weeks of the latest month to reflect the fact that, in the out-of-sample simulations, we time production of our nowcasts and estimates to be coincident with BEA releases of U.S. GDP that typically happen in the final week of a given month.

Given that real state GDP data, at the annual frequency, are available only from 1977 we obtain annual estimates of real state GDP back to 1963 (1964 in annual growth rates) by deflating the available nominal state GDP data by the U.S. GDP deflator. Given that the BLS does not publish state-level price indices, \cite{del2002asymmetric} also used the U.S. GDP deflator to produce real state GDP data, finding that the use of the U.S. deflator produced similar estimates of real GDP to use of CPI data when available by state. \cite{NakamuraQJE} recently construct new state-level prices indices for the U.S. states using BLS micro data, but these date back to 1978 meaning they cannot help us deflate the nominal state GDP data available from 1963 through 1977. For this reason, we use the annual GDP by state (series id: SAGDP2) and the U.S. GDP deflator (series id: Gross Domestic Product: Implicit Price Deflator (A191RI1A225NBEA)). 

Personal income (XXOTOT) and wage and salary (SQINC5N) data, as listed in Table \ref{Tab:Data}, are only available in nominal terms at the state-level. We convert them to real variables by deflating using the U.S. deflator. \cite{arias2016metro} take a similar approach. For both of these variables, data are available for some states back to the beginning of our sample (1964m1) but not for all states. Typically coverage starts, for those missing states, in the 1970s.

To estimate the unemployment rate for  state XX, (XXUR), from 1964 to 1975, we backcast based on national unemployment trends. While official data for state XX are only available from 1976 onward, we extend the time series back by using the growth rate of the U.S. unemployment rate from 1964 to 1975. This approach assumes that the state's unemployment patterns were closely aligned with national trends during these years. By applying the year-over-year growth of the U.S. unemployment rate to the 1976 value for state XX, we generate a consistent time series that fills in the gaps in the data, allowing for historical analysis of the state’s labor market prior to the availability of direct statistics.

To backcast the initial claims series for state XX (XXICLAIMS) from 1964 to 1985, we employ an ordinary least squares (OLS) regression analysis that uses both state and national unemployment claims data. The initial claims series for state XX begins in 1986, but to extend this series backward, we first regress the state claims against national claims data, including a constant term. This regression will identify the relationship between state and national trends in unemployment claims. Once we have the regression coefficients, we can apply them to the U.S. initial claims series for the years 1964 to 1985. By substituting these coefficients into the equation, we can estimate the state’s initial claims during this earlier period, effectively creating a consistent dataset that reflects both state-specific dynamics and national trends leading up to 1986.

For some variables, real-time data vintages were occasionally missing at specific points in time. For example, the 01/25/2019 vintage of U.S. GDP is missing due to the partial government shutdown. Our general strategy is to fill in occasional missing vintages, for some variables, with (from the perspective of that data vintage) the most recently available data vintage and then forecast any missing observation via a random walk. 

\begin{landscape}
    \begin{table}[]
        \centering
        \caption{The column tcode denotes the following data transformation for a series x: (1) no transformation; (2) $\Delta$xt ; (3) $\Delta^2$xt ; (4) $\log$ (xt); (5) $\Delta$ $\log$(xt); (6) $\Delta^2$ $\log$(xt); (7) $\Delta^2$ $(x_t/x_{t-1}-1)$; (8)  $(x_t/x_{t-1}-1)$.The FRED column gives mnemonics in FRED followed by a short description. Seasonally unadjusted series seasonally adjusted using X11. XX denotes the short code for U.S. state. n denotes that the data were available only in nominal terms, r denotes real, and * denotes that the data were not available back to this date for all states.}
        \label{t5}
        \resizebox{22cm}{!}{%
            \begin{tabular}{@{}llllllllllc@{}}  
                \toprule
                & tcode & mnemonics                & Description                                & Category  & SA & Frequency & First obs & Source & \multicolumn{2}{l}{First vintage} \\ \midrule
                1 & 8   & GDPC1                  & Real Gross Domestic Product                        & U.S. level  & Yes         & Quarterly & 1964Q1    & ALFRED    & \multicolumn{2}{l}{21/12/2006}   \\
                2 & 6   & FPIC1                  & Fixed Investment                             & U.S. level  & Yes         & Quarterly & 1964Q1    & ALFRED    & \multicolumn{2}{l}{21/12/2006}   \\
                3 & 5   & GCEC1                  & Government Expenditures                          & U.S. level  & Yes         & Quarterly & 1964Q1    & ALFRED    & \multicolumn{2}{l}{21/12/2006}   \\
                4 & 2   & UNRATE                 & Civilian Unemployment Rate                        & U.S. level  & Yes         & Monthly  & 1964M1    & ALFRED    & \multicolumn{2}{l}{08/12/2006}   \\
                5 & 6   & CPIAUCSL                & CPI : All Items                              & U.S. level  & Yes         & Monthly  & 1964M1    & ALFRED    & \multicolumn{2}{l}{15/12/2006}   \\
                6 & 5   & INDPRO                 & IP Index                                 & U.S. level  & Yes         & Monthly  & 1964M1    & ALFRED    & \multicolumn{2}{l}{15/12/2006}   \\
                7 & 5   & PCEC96                 & Personal consumption expenditure                     & U.S. level  & Yes         & Monthly  & 1964M1    & ALFRED    & \multicolumn{2}{l}{22/12/2006}   \\
                8 & 2   & FEDFUNDS                & Effective Federal Funds Rate                       & U.S. level  & No         & Monthly  & 1964M1    & ALFRED    & \multicolumn{2}{l}{04/12/2006}   \\
                9 & 2   & GS10                  & 10-Year Treasury Rate                           & U.S. level  & No         & Monthly  & 1964M1    & ALFRED    & \multicolumn{2}{l}{04/12/2006}   \\
                10 & 5   & SP500                  & S\&P Common Stock Price Index: Composite                & U.S. level  & No         & Monthly  & 1964M1    & \multicolumn{2}{l}{Bloomberg} & -        \\
                11 & 2   & AWHMAN                  & Avg Weekly Hours of Prod. and Nonsupervisory Employees, Manufacturing
                 & U.S. level  & Yes         & Monthly  & 1964M1    & ALFRED    & \multicolumn{2}{l}{08/12/2006}   \\
                12 & 5   & RPI                   & Real personal income                           & U.S. level  & Yes         & Monthly  & 1964M1(n)    & ALFRED    & \multicolumn{2}{l}{22/12/2006}   \\
                13 & 5   & PAYEMS                 & All Employees: Total nonfarm                       & U.S. level  & Yes         & Monthly  & 1964M1    & ALFRED    & \multicolumn{2}{l}{08/12/2006}   \\
                14 & 6   & OILPRICE, MCOILWTICO          & Producer price index, crude petroleum, 1982=1.0              & U.S. level  & No         & Monthly  & 1964M1    & ALFRED    & \multicolumn{2}{l}{01/12/2006}   \\
                15 & 8   & SAGDP9                 & Real GDP by state                             & State level & No         & Annual  & 1964(n) 1977(r)     & BEA     & \multicolumn{2}{l}{07/06/2007}   \\
                16 & 8   & SQGDP9                 & Real GDP by state                             & State level & Yes         & Quarterly & 2005Q1    & BEA     & \multicolumn{2}{l}{27/07/2016}   \\
                17 & 5   & XXOTOT                 & Real personal income                           & State level & Yes         & Quarterly & 1964Q1*    & ALFRED    & \multicolumn{2}{l}{20/09/2010}   \\
                18 & 5   & SQINC5N                 & Wage and salary                              & State level & No         & Quarterly & 1964Q1(n)    & BEA     & \multicolumn{2}{c}{-}       \\
                19 & 5   & XXICLAIMS                & Initial Claims                              & State level & No         & Monthly  & 1986M1    & ALFRED    & \multicolumn{2}{l}{09/09/2010}   \\
                20 & 5   & XXNAN                  & All Employees: Total Nonfarm                       & State level & No         & Monthly  & 1964M1    & ALFRED    & \multicolumn{2}{l}{19/06/2007}   \\
                21 & 2   & SMUXX000000000000007, SAUXX000000000005 & Average Weekly Hours: Manuf        & State level & No         & Monthly  & 1964M1*    & BLS     & \multicolumn{2}{c}{-}       \\
                22 & 2   & XXUR                  & Unemployment Rate                             & State level & Yes         & Monthly  & 1976M1    & ALFRED    & \multicolumn{2}{l}{19/06/2007}   \\ \bottomrule
            \end{tabular}%
        }  \label{Tab:Data}
    \end{table}
\end{landscape}

\begin{sidewaystable}[]
    \centering
    \caption{Illustrative Data Release Calendar for Nowcasts/Estimates Made at the End of the Month Indicated}
    \label{t3}
    \resizebox{\textwidth}{!}{%
        \begin{tabular}{lcccccccccccc}
            \hline
            & Jan   & Feb   & Mar   & Apr   & May   & Jun   & Jul   & Aug   & Sep   & Oct   & Nov   & Dec   \\ \hline
            Real GDP     & Q4     & $Q4^*$     & $Q4^{**}$    & Q1     & $Q1^*$     & $Q1^{**}$     & Q2     & $Q2^*$     & $Q2^{**}$     & Q3     & $Q3^*$     & $Q3^{**}$     \\
            Fixed Inv     & Q4     & $Q4^*$     & $Q4^{**}$     & Q1     & $Q1^*$     & $Q1^{**}$     & Q2     & $Q2^*$     & $Q2^{**}$     & Q3     & $Q3^*$     & $Q3^{**}$     \\
            Gov Exp      & Q4     & $Q4^*$     & Q4     & Q1     & Q1     & Q1     & Q2     & Q2     & Q2     & Q3     & Q3     & Q3     \\
            UNM        & M12    & M1     & M2     & M3     & M4     & M5     & M6     & M7     & M8     & M9     & M10     & M11     \\
            CPI        & M12    & M1     & M2     & M3     & M4     & M5     & M6     & M7     & M8     & M9     & M10     & M11     \\
            IP        & M12    & M1     & M2     & M3     & M4     & M5     & M6     & M7     & M8     & M9     & M10     & M11     \\
            EFFR       & M1(3)     & M2(3)     & M3(3)     & M4(3)     & M5(3)     & M6(3)     & M7(3)     & M8(3)     & M9(3)      & M10(3)      & M11(3)      & M12(3)  \\
            10 T-bill     & M1(3)     & M2(3)     & M3(3)     & M4(3)     & M5(3)     & M6(3)     & M7(3)     & M8(3)     & M9(3)      & M10(3)      & M11(3)      & M12(3)  \\
            SnP500      & M1(3)     & M2(3)     & M3(3)     & M4(3)     & M5(3)     & M6(3)     & M7(3)     & M8(3)     & M9(3)      & M10(3)      & M11(3)      & M12(3)      \\
            AVG hrs worked  & M12    & M1     & M2     & M3     & M4     & M5     & M6     & M7     & M8     & M9     & M10     & M11     \\
            EMP        & M12    & M1     & M2     & M3     & M4     & M5     & M6     & M7     & M8     & M9     & M10     & M11     \\
            RPI        & M12    & M1     & M2     & M3     & M4     & M5     & M6     & M7     & M8     & M9     & M10     & M11     \\
            Real GDP by state & Q3     & Q3     & Q4     & Q4     & Q4     & Q1     & Q1     & Q1     & Q2     & Q2     & Q2     & Q3     \\
            Real GDP by state & $A_{t-2}$ & $A_{t-2}$ & $A_{t-1}$ & $A_{t-1}$ & $A_{t-1}$ & $A_{t-1}$ & $A_{t-1}$ & $A_{t-1}$ &$A_{t-1}$ & $A_{t-1}$ & $A_{t-1}$ & $A_{t-1}$ \\
            RPI      & Q3     & Q3     & Q4     & Q4     & Q4     & Q1     & Q1     & Q1     & Q2     & Q2     & Q2     & Q3     \\
            Wages Salaries  &Q3     & Q3     & Q4     & Q4     & Q4     & Q1     & Q1     & Q1     & Q2     & Q2     & Q2     & Q3     \\
            Initial Claims  &  M1(3)     & M2(3)     & M3(3)     & M4(3)     & M5(3)     & M6(3)     & M7(3)     & M8(3)     & M9(3)      & M10(3)      & M11(3)      & M12(3)      \\
            EMP        & M12    & M1     & M2     & M3     & M4     & M5     & M6     & M7     & M8     & M9     & M10     & M11     \\
            AVG hrs worked  & M12    & M1     & M2     & M3     & M4     & M5     & M6     & M7     & M8     & M9     & M10     & M11      \\
            UNM        & M12    & M1     & M2     & M3     & M4     & M5     & M6     & M7     & M8     & M9     & M10     & M11     \\ \hline
        \end{tabular}%
    }
\footnotesize{Notes: For Wages Salaries and AVG hrs worked series, data are not revised. ``*'' and ``**'' denote the second and third data-releases. ``3'' in brackets means that the monthly average is computed using the first 3 weeks of data for the month indicated, given that our nowcasts are timed as being produced in the last week of each month, before the final week's data for the month is available.} 
\end{sidewaystable}

\clearpage
\newpage
\section{Technical Appendix}\label{TechAppendix}

The MCMC algorithm used in this paper involves drawing latent states (the high frequency values of the variables which are only observed at a low frequency) conditional on the remaining parameters and drawing the parameters conditional on the draws of the latent states. We will describe these two blocks of the algorithm in separate sub-sections. 

\subsection{Drawing the Parameters of the MF-VAR}
Conditional on the latent states, we simply have a VAR and standard methods for drawing the paramters of the VAR when using the horseshoe prior can be used. These are described in this sub-section. 
 
\begin{enumerate}
    \item The conditional posterior distribution of the regression coefficients $\theta_{i}$ is:
    \begin{equation}
        \theta_{i} \sim N\left( A_{i}^{-1}X_i^{'} y_i,  \sigma_i^2 A_{i}^{-1} \right), 
    \end{equation}%
    where $A _{i}= \left( X_i^{'}X_i + \Lambda_{\star}^{-1} \right)$, $\Lambda_{\star} = \tau_i^2 \Lambda_i^2$. We use the efficient algorithm of \cite{bhattacharya2016fast}, which makes use of the Woodbury identity, to sample them.
    
    \item The conditional posterior distribution of $\sigma^2_i$ is:
    
    \begin{equation}
        \sigma_i^2 \sim IG \left( \bar{ \nu_{\sigma}} , \bar{S_{\sigma}} \right), 
    \end{equation}%
    where $\bar{ \nu_{\sigma}} = (T+k_i)/2$ and $\bar{S_{\sigma}} = \left( y_i - X_i\theta_{i}  \right)^{'}  \left( y_i - X_i\theta_{i}  \right)/2 + \theta_{i}^{'} \Lambda_{\star}^{-1}\theta_{i}/2$
    \item The conditional posterior distribution of $\tau^2_i$ can be drawn using slice sampling as described in the Appendix of \cite{korobilis2022new}.

        \item The conditional posterior distribution of $\lambda^2_{i,j}$ can be drawn using slice sampling as described in the Appendix of \cite{korobilis2022new}.

\end{enumerate}

\subsection{Drawing the States of the MF-VAR}
Remember that $y_{t}$ is the N-dimensional series, in which not all its variables will be observed every period under a mixed frequency setup. We consider $y_t = (y_{m, t}^{US},y_{q,t}^{US}, y_{a}^{S})^{'}$, where $N_m$ collects the monthly U.S. variables, $y_{m, t}^{US}$, such as inflation, $N_q$ shows the quarterly U.S. and state variables post-2005, $y_{q,t}^{US}$,$y_{q,t}^{S}$, respectively that are observed every quarter and finally $N_a$ collects the annual GDP by state, which is observed every year pre-2005. It holds that $N = N_m + N_q+ N_a$. 
    
To describe the monthly and quarterly dynamics, we assume $y_{q,m,t}$, $y_{a,m,t}$, $y_{a,q,t}$ denote the monthly and quarterly latent variables underlying the quarterly series and annual series, $y_{q,t}^{US}$,$y_{q,t}^{S}$, and $y_{a}^{S}$, respectively. As described in the body of the paper, we run our MCMC in two steps; the former considers the annual --quarterly mismatch, while the latter depicts the monthly -- quarterly frequency. We combine these latent variables with the indicators observed at a quarterly and a monthly frequency in $X^{k}_t= [X^{HF}_t, X^{LF}_t]$, $k=1,2$\footnote{1 corresponds to quarterly indicators, while 2 depicts the monthly latent variables, and HF stands for high frequency data and LF for low frequency}. Clearly, when $k=1$, HF is the quarterly data and LF is the annual variables; while, when $k=2$, HF is the monthly indicator and LF is the quarterly series.

To alleviate the computational burden of the state space MF-VAR, we follow \cite{schorfheide2015real} and \cite{ankargren2021simulation} which imply that the monthly variables can be omitted from the state equation and instead enter the system through the exogenous terms in a state-space model. However, this implies that the state and measurement errors will be correlated, which thus requires the use of an algorithm based on a state-space model with between-equation correlation. The companion form of the monthly VAR together with a measurement equation for $y_t$ delivers the common two-equation state-space system given by:
\begin{align}\label{mfvarss}
    y_t &= C_t + Z_t S_t + G_t \epsilon_{t}\\ \label{mfvarss2}
    S_t &= D_t + T_t S_{t-1} + H_t \epsilon_t \\ 
    \epsilon_{t} &\sim N\left(0, I_N\right),
\end{align}
where:
\begin{align}
    Z_t &= \begin{pmatrix}
        0_{N_{HF} \times N_{LF}} & \Phi_{HF, LF} \\
        \Lambda_{LF} & 0_{N_{LF} \times N_{LF}}
    \end{pmatrix}, \label{selectionM} \\
    T_t &= \begin{pmatrix}
        \Phi_{LF, LF} & 0_{N_{LF} \times N_{LF}} \\
        I_{pN_{LF}} & 0_{pN_{LF} \times N_{LF}}
    \end{pmatrix}, \\
C_t &= \begin{pmatrix}
    \Phi_{HF, HF}& \Phi_{HF, 0} \\
    0_{N_{LF} \times N_{HF}}& 0_{N_{LF} \times 1}
\end{pmatrix}\begin{pmatrix}
y_{m,t-1:t-p} \\
1
\end{pmatrix} + 
\begin{pmatrix}
\beta X_{HF,t-1:t-p} \\
0_{N_{LF} \times 1}
\end{pmatrix}, \\
D_t &= \begin{pmatrix}
    \Phi_{LF, HF}& \Phi_{LF, 0} \\
    0_{pN_{LF} \times pN_{HF}} & 0_{pN_{LF} \times 1}
\end{pmatrix}\begin{pmatrix}
    y_{m,t-1:t-p} \\
    1
\end{pmatrix}, \\
G_t &= \begin{pmatrix}
    \Sigma^{1/2}_{HF}\\
    0_{N_{LF} \times N}
\end{pmatrix}, \\
H_t &= \begin{pmatrix}
    \Sigma^{1/2}_{LF}\\
    0_{pN_{LF} \times N}
\end{pmatrix}, \\
\Sigma_t^{1/2} &= \begin{pmatrix}
    \Sigma^{1/2}_{HF}\\
    \Sigma^{1/2}_{LF}
\end{pmatrix}. 
\end{align}
We present the companion form of the VAR at the high frequency as:
\begin{align}
 \begin{pmatrix}
    X_{t}^k\\
    X_{t-1}^k\\
    \vdots \\
    X_{t-p+1}^k\\
\end{pmatrix} &= 
 \begin{pmatrix}
    \Phi_{0}\\
    0\\
    \vdots \\
    0\\
\end{pmatrix} + 
 \begin{pmatrix}
    \Phi_1 & \Phi_2 & \ldots &\Phi_{p-1} &\Phi_{p} \\
    I_N &0 &\ldots &0 &0\\
    \vdots & \vdots & \ddots & \vdots & \vdots\\
    0 &0 & \ldots &I_N &0\\
\end{pmatrix} \begin{pmatrix}
X_{t-1}^k\\
X_{t-2}^k\\
\vdots \\
X_{t-p}^k\\
\end{pmatrix} + 
 \begin{pmatrix}
    u_t\\
    0\\
    \vdots \\
    0\\
\end{pmatrix} \\
S_t &= \begin{pmatrix}
    X_{t}^k\\
    X_{t-1}^k\\
    \vdots \\
    X_{t-p+1}^k\\
\end{pmatrix} , \Phi_0 = \begin{pmatrix}
\Phi_{HF,0}\\
\Phi_{LF,0}
\end{pmatrix} , 
\Phi_i = \begin{pmatrix}
    \Phi_{HF,HF} & \Phi_{HF,LF}\\
    \Phi_{LF, HF} & \Phi_{LF, LF}
\end{pmatrix}.
\end{align}

Finally, in (\ref{selectionM}) we augment $Z_t$ with the cross sectional restriction. $\beta X_{HF,t-1:t-p} $ collects the monthly state-level indicators.

We follow algorithm 2 of \cite{durbin2002simple} to generate draws from $P(S|\Theta, Y)$. Finally, when smoothing, $y_t^{\star}$ we set constant terms equal to zero, as in \cite{jarocinski2015note}.

\clearpage

\newpage
\clearpage 

\section{Empirical Appendix} \label{EmpiricsAppendix}    
In the main body of this paper we work with exact growth rates. In this Appendix, we first show that using log differences instead makes little difference to our results and the monthly state GDP estimates look similar.

\begin{figure}[!h]
  \caption{Historical monthly estimates of state-level GDP growth using log differences expressed as year-on-year growth rates ($y^s_{a,t}$), for nine selected states}
  \centering
\includegraphics[scale=0.38]{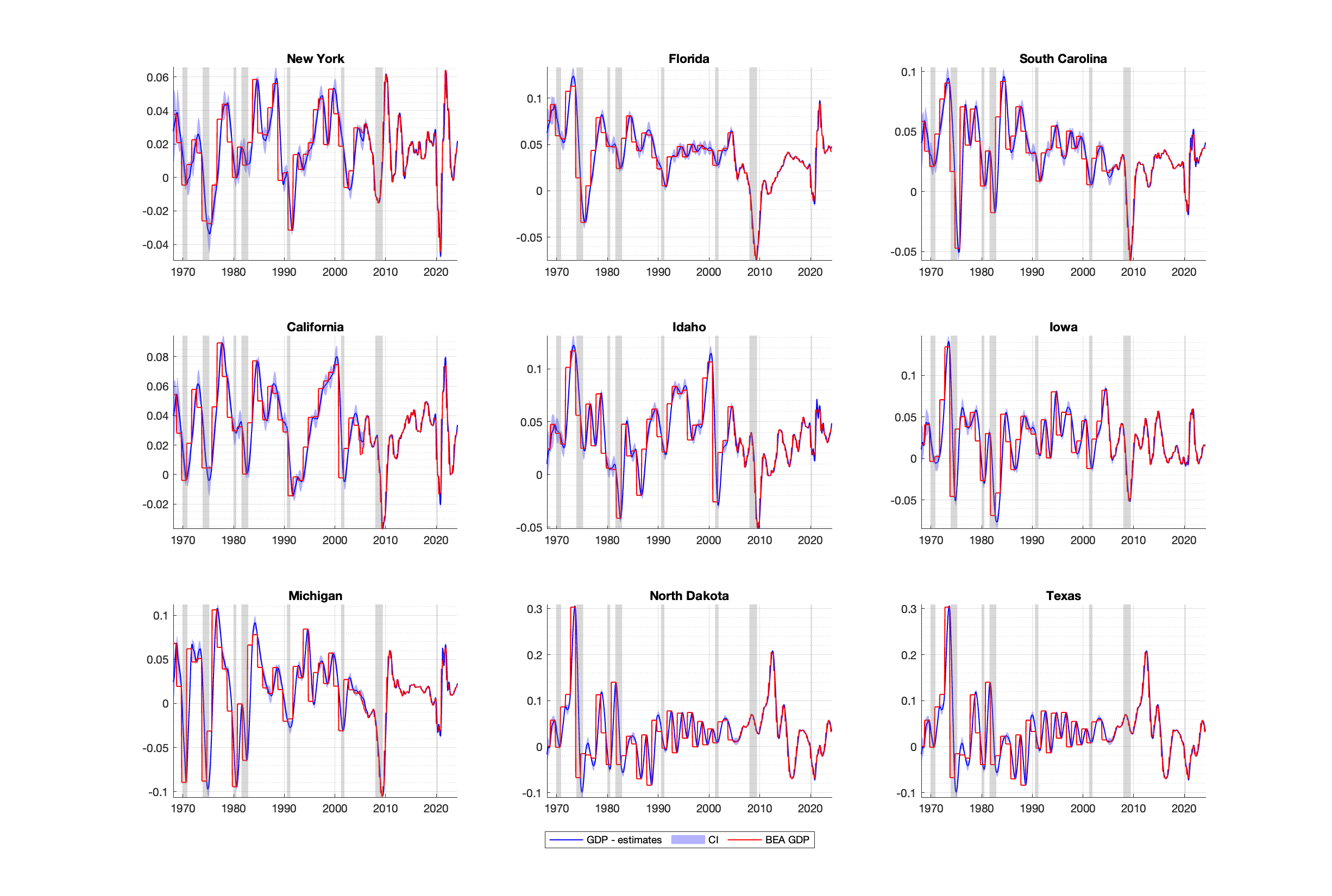}
  
    \label{fig:sreglog}
    \footnotesize{Notes: NBER recession bands in gray. 68 percent credible intervals around the posterior median estimates shown in blue. Red lines denote BEA state-level GDP growth rate, available at the annual frequency prior to 2005, quarterly thereafter.}
\end{figure}

\begin{table}[]
\caption{Average Log Predictive Scores (ALPS) for quarterly GDP estimates and nowcasts, 2007Q1-2024Q1}
\label{ALPStb1}
\centering
\resizebox{.93\columnwidth}{!}{%
\begin{tabular}{@{}lccccc@{}}
\toprule
                       & m1 nowcast $\tau$ & m2 nowcast $\tau$ & m3 nowcast $\tau$ & m1 estimate $\tau-1$ & m2 estimate $\tau-1$ \\ \midrule
U.S.                    & 263.90            & 263.43            & 315.43            & --                 & --                \\
Alabama                & 283.39            & 283.34            & 299.98            & 300.39               & 300.30               \\
Alaska                 & 232.93            & 231.52            & 239.18            & 241.88               & 238.45               \\
Arizona                & 237.50            & 237.16            & 256.74            & 258.34               & 257.16               \\
Arkansas               & 264.46            & 266.48            & 293.39            & 293.84               & 294.87               \\
California             & 234.17            & 234.26            & 261.41            & 261.39               & 265.66               \\
Colorado               & 272.94            & 273.07            & 289.91            & 292.31               & 291.01               \\
Connecticut            & 247.50            & 246.76            & 262.03            & 265.00               & 267.18               \\
Delaware               & 218.57            & 217.48            & 229.54            & 228.72               & 226.31               \\
District of Columbia & 297.64            & 298.08            & 301.32            & 301.39               & 301.09               \\
Florida                & 249.04            & 248.50            & 282.15            & 284.11               & 284.41               \\
Georgia                & 249.39            & 248.99            & 280.13            & 281.72               & 283.55               \\
Hawaii                 & 278.66            & 277.94            & 289.17            & 291.32               & 294.15               \\
Idaho                  & 274.61            & 273.98            & 284.96            & 284.65               & 284.36               \\
Illinois               & 284.93            & 284.67            & 301.40            & 302.15               & 304.34               \\
Indiana                & 271.53            & 272.78            & 287.55            & 288.60               & 289.45               \\
Iowa                   & 280.65            & 279.87            & 286.29            & 288.65               & 288.34               \\
Kansas                 & 270.19            & 270.69            & 283.75            & 285.07               & 284.84               \\
Kentucky               & 282.36            & 282.54            & 289.94            & 290.53               & 290.47               \\
Louisiana              & 265.95            & 265.81            & 279.13            & 279.25               & 280.77               \\
Maine                  & 290.21            & 290.03            & 299.16            & 299.44               & 298.35               \\
Maryland               & 265.67            & 265.26            & 285.77            & 286.27               & 289.05               \\
Massachusetts          & 276.73            & 277.58            & 291.97            & 292.32               & 291.73               \\
Michigan               & 271.31            & 272.48            & 280.86            & 285.54               & 286.31               \\
Minnesota              & 276.59            & 277.04            & 292.59            & 293.52               & 294.15               \\
Mississippi            & 277.98            & 277.88            & 290.92            & 290.89               & 290.18               \\
Missouri               & 283.38            & 284.53            & 298.01            & 298.13               & 297.88               \\
Montana                & 283.10            & 281.87            & 288.65            & 289.00               & 288.92               \\
Nebraska               & 271.74            & 270.62            & 264.33            & 266.75               & 267.88               \\
Nevada                 & 224.53            & 222.92            & 260.06            & 264.91               & 265.39               \\
New Hampshire         & 227.67            & 226.44            & 228.41            & 233.57               & 233.41               \\
New Jersey            & 276.51            & 276.91            & 293.70            & 294.11               & 294.11               \\
New Mexico            & 247.14            & 248.30            & 266.99            & 266.93               & 267.21               \\
New York              & 250.55            & 250.78            & 268.49            & 267.59               & 268.72               \\
North Carolina        & 259.30            & 259.47            & 289.81            & 289.05               & 289.43               \\
North Dakota          & 235.25            & 235.16            & 242.03            & 244.72               & 244.32               \\
Ohio                   & 289.58            & 289.99            & 299.85            & 300.87               & 301.35               \\
Oklahoma               & 273.12            & 274.11            & 274.17            & 274.50               & 274.26               \\
Oregon                 & 255.98            & 256.30            & 278.25            & 276.22               & 276.40               \\
Pennsylvania           & 286.80            & 286.35            & 298.65            & 298.77               & 299.79               \\
Rhode Island          & 273.29            & 272.16            & 281.73            & 283.03               & 283.33               \\
South Carolina        & 269.65            & 269.42            & 289.40            & 291.34               & 292.76               \\
South Dakota          & 237.21            & 236.67            & 222.97            & 225.39               & 224.43               \\
Tennessee              & 275.99            & 276.26            & 295.26            & 294.35               & 293.72               \\
Texas                  & 244.79            & 245.27            & 271.26            & 273.96               & 276.99               \\
Utah                   & 265.24            & 264.95            & 282.06            & 283.25               & 283.80               \\
Vermont                & 271.25            & 269.60            & 285.32            & 285.68               & 283.78               \\
Virginia               & 261.72            & 260.58            & 290.23            & 290.25               & 291.07               \\
Washington             & 265.01            & 265.06            & 276.83            & 279.45               & 280.49               \\
West Virginia         & 286.07            & 286.62            & 286.49            & 285.87               & 285.87               \\
Wisconsin              & 282.52            & 282.27            & 298.86            & 299.94               & 301.98               \\
Wyoming                & 255.91            & 256.40            & 251.41            & 254.01               & 252.88               \\ \bottomrule
AVG                    & 264.87            & 264.77            & 278.87            & 279.98               & 280.33                     \\ \bottomrule
\end{tabular}
}
\footnotesize{Notes: ALPS statistics x 100. AVG denotes the equal-weighted average across all states. 2020Q2-Q4 dropped to avoid contamination due to COVID outliers.}
\end{table}

\begin{table}[]
\caption{Average Log Predictive Scores relative to benchmark state-specific MF-VAR}
\resizebox{.92\columnwidth}{!}{%
\begin{tabular}{@{}lccccc@{}}
\toprule
                       & m1 nowcast $\tau$ & m2 nowcast $\tau$ & m3 nowcast $\tau$ & m1 estimate $\tau-1$ & m2 estimate $\tau-1$ \\ \midrule
U.S.                    & 0.31              & 0.32              & 0.32              &  --                    &   --                   \\
Alabama                & 0.11              & 0.12              & 0.01              & -0.06                & -0.04                \\
Alaska                 & 0.14              & 0.11              & 0.56              & 0.55                 & 0.55                 \\
Arizona                & 0.00              & 0.01              & 0.34              & 0.27                 & 0.29                 \\
Arkansas               & 0.04              & 0.05              & 0.02              & -0.04                & -0.04                \\
California             & -0.41             & -0.40             & -0.18             & -0.21                & -0.18                \\
Colorado               & -0.07             & -0.06             & -0.07             & -0.08                & -0.06                \\
Connecticut            & 0.08              & 0.09              & 0.21              & 0.15                 & 0.15                 \\
Delaware               & 0.01              & 0.03              & 0.21              & 0.15                 & 0.18                 \\
District of Columbia & -0.13             & -0.13             & -0.02             & -0.03                & -0.02                \\
Florida                & -0.20             & -0.19             & -0.09             & -0.13                & -0.11                \\
Georgia                & -0.03             & -0.02             & 0.03              & -0.04                & -0.01                \\
Hawaii                 & -0.13             & -0.11             & -0.10             & -0.13                & -0.14                \\
Idaho                  & -0.03             & -0.03             & 0.17              & 0.16                 & 0.17                 \\
Illinois               & 0.01              & 0.01              & 0.04              & -0.01                & -0.01                \\
Indiana                & 0.07              & 0.08              & 0.22              & 0.17                 & 0.18                 \\
Iowa                   & 0.04              & 0.05              & 0.23              & 0.17                 & 0.20                 \\
Kansas                 & 0.08              & 0.07              & 0.14              & 0.10                 & 0.13                 \\
Kentucky               & 0.12              & 0.13              & 0.24              & 0.18                 & 0.20                 \\
Louisiana              & 0.06              & 0.06              & 0.05              & 0.00                 & -0.01                \\
Maine                  & 0.08              & 0.09              & 0.19              & 0.15                 & 0.18                 \\
Maryland               & 0.06              & 0.06              & 0.07              & 0.04                 & 0.04                 \\
Massachusetts          & -0.06             & -0.07             & -0.04             & -0.05                & -0.03                \\
Michigan               & -0.07             & -0.06             & 0.02              & 0.00                 & 0.03                 \\
Minnesota              & 0.04              & 0.02              & 0.22              & 0.19                 & 0.19                 \\
Mississippi            & 0.07              & 0.09              & 0.17              & 0.11                 & 0.15                 \\
Missouri               & 0.05              & 0.03              & 0.05              & 0.03                 & 0.04                 \\
Montana                & 0.08              & 0.10              & 0.23              & 0.20                 & 0.20                 \\
Nebraska               & 0.11              & 0.11              & 0.46              & 0.40                 & 0.40                 \\
Nevada                 & -0.20             & -0.19             & -0.03             & -0.08                & -0.07                \\
New Hampshire         & 0.01              & 0.03              & 0.52              & 0.57                 & 0.54                 \\
New Jersey            & -0.06             & -0.05             & 0.05              & 0.04                 & 0.06                 \\
New Mexico            & 0.02              & 0.03              & 0.07              & 0.07                 & 0.08                 \\
New York              & -0.10             & -0.12             & 0.23              & 0.18                 & 0.19                 \\
North Carolina        & 0.08              & 0.08              & 0.14              & 0.12                 & 0.14                 \\
North Dakota          & 0.17              & 0.18              & 0.48              & 0.46                 & 0.46                 \\
Ohio                   & -0.04             & -0.03             & -0.02             & -0.08                & -0.04                \\
Oklahoma               & -0.08             & -0.08             & 0.09              & 0.07                 & 0.08                 \\
Oregon                 & 0.01              & 0.02              & 0.22              & 0.18                 & 0.19                 \\
Pennsylvania           & 0.02              & 0.02              & -0.01             & -0.03                & -0.04                \\
Rhode Island          & 0.04              & 0.04              & 0.28              & 0.22                 & 0.21                 \\
South Carolina        & 0.08              & 0.06              & 0.10              & 0.03                 & 0.04                 \\
South Dakota          & 0.17              & 0.17              & 0.92              & 0.82                 & 0.86                 \\
Tennessee              & 0.06              & 0.08              & 0.10              & 0.06                 & 0.10                 \\
Texas                  & -0.11             & -0.09             & -0.08             & -0.10                & -0.09                \\
Utah                   & 0.07              & 0.06              & 0.06              & 0.03                 & 0.03                 \\
Vermont                & 0.03              & 0.04              & 0.16              & 0.13                 & 0.17                 \\
Virginia               & 0.14              & 0.14              & 0.03              & -0.01                & -0.01                \\
Washington             & -0.05             & -0.05             & 0.02              & -0.01                & -0.01                \\
West Virginia         & 0.02              & 0.03              & 0.16              & 0.16                 & 0.18                 \\
Wisconsin              & 0.09              & 0.10              & 0.08              & 0.02                 & 0.03                 \\
Wyoming                & 0.10              & 0.10              & 0.32              & 0.29                 & 0.31                 \\ \bottomrule
AVG                    & 0.01              & 0.02              & 0.14              & 0.11                 & 0.12     \\ \bottomrule
\end{tabular}

}\\
\footnotesize{Notes: Estimates computed as log predictive score of MF-VAR model, (\ref{VAR}), minus log predictive score of benchmark state-specific MF-VAR. Values greater than zero indicate superior performance of the MF-VAR model, (\ref{VAR}). 2020Q2-Q4 dropped to avoid contamination due to COVID outliers.}
\end{table}

\end{document}